\documentclass[twocolumn]{aastex63} 
\usepackage{graphicx}	
\usepackage{amsmath}	
\usepackage{amssymb}	
\usepackage{physics}
\usepackage{enumitem}
\usepackage{bbm}
\usepackage{color}
\usepackage{natbib}
\newcommand\msun{$M\sb{\odot}$}

\newcommand{\AV}{$A_V$}

\newcommand{\fwise}{f_\mathrm{12}/f_\mathrm{4.6}}
\newcommand{\lfwise}{ \log\, (f_\mathrm{12}/f_\mathrm{4.6})}

\newdimen\digitwidth      
\setbox1=\hbox{0}       
\digitwidth=\wd1        
\catcode`"=\active      
\def"{\kern\digitwidth}

\begin{document}

\title{Some Die Filthy Rich: The Diverse Molecular Gas Contents of Post-starburst Galaxies Probed by Dust Absorption}
\shorttitle{Dust Absorption and Molecular Gas in PSBs}
\shortauthors{Yesuf \& Ho}
\author{Hassen M. Yesuf}
\affiliation{Kavli Institute for Astronomy and Astrophysics, Peking University, Beijing 100871, China}
\affiliation{Kavli Institute for the Physics and Mathematics of the Universe, The University of Tokyo, Kashiwa, Japan 277-8583}
\author{Luis C. Ho}
\affiliation{Kavli Institute for Astronomy and Astrophysics, Peking University, Beijing 100871, China}
\affiliation{Department of Astronomy, School of Physics, Peking University, 
Beijing 100871, China}

\begin{abstract}
Quenched post-starburst galaxies (QPSBs) are a rare but important class of galaxies that show signs of rapid cessation or recent rejuvenation of star formation. A recent observation shows that about half of QPSBs have large amounts of cold gas. This molecular CO sample is, however, too small and is not without limitations. Our work aims to verify previous results by applying a new method to study a uniformly selected sample, more than 10 times larger. In particular, we present detailed analysis of H$\alpha$/H$\beta$ ratios of face-on QPSBs at $z = 0.02 - 0.15$ and with $M_\star = 10^{10}-10^{11}\,M_\odot$. We interpret the H$\alpha$/H$\beta$ ratios by applying our recent gas mass calibration, which is based on non-PSB galaxies but predicts gas masses that are consistent with CO observations of $\sim 100$ PSBs. We estimate the molecular gas by either using PSBs with well-measured H$\alpha$/H$\beta$ ratios or by measuring them from stacked spectra. Our analysis reveals that QPSBs have a wide range of H$\alpha$/H$\beta$ ratios and molecular gas fractions that overlap with the typical gas fractions of star-forming or quiescent galaxies: H$\alpha$/H$\beta \approx 3-8$ and $f_\mathrm{H_2} \approx 1\%-20\%$ with median $f_\mathrm{H_2} \approx 4\%-6\%$, which correspond to $M_\mathrm{H_2} \approx (1-3) \times 10^{9} \,M_\odot$. Our results indicate that large reservoirs of cold gas are still present in significant numbers of QPSBs and that they arguably were not removed or destroyed by feedback from active galactic nuclei.
\end{abstract}


\section{INTRODUCTION}

Massive galaxies in the nearby universe fall into two broad categories: star-forming, gas-rich spiral galaxies and quiescent, gas-poor lenticular and elliptical galaxies. Understanding the origin of this bimodality has been the subject of considerable research for the past several decades. In the traditional picture, post-starburst (PSB) galaxies \citep{Dressler+83, Quintero+04, Goto+05, Wild+09, Yesuf+14,Pawlik+18} may exemplify objects in transition from the star-forming to quiescent phase owing to major mergers \citep{Snyder+11}. Minor mergers, however, may equally produce a large fraction of rejuvenated PSBs, by triggering new cycles of starbursts in passive, bulge-dominated galaxies \citep{Dressler+13,Rowlands+18,Davis+19,Pawlik+19}. In their quiescent phase, PSB spectra reveal very little ongoing star formation (weak emission lines) but a substantial burst of star formation (strong Balmer absorption lines) in their recent past ($\lesssim 1$ Gyr). Although PSBs are rare galaxies ($\lesssim 1\%$), observational estimates of their rapid evolution timescale and number density indicate that a significant fraction of present-day ellipticals may have gone through the PSB evolutionary channel \citep[e.g.,][]{Tran+04,Yesuf+14, Wild+16,Zahid+16}. Several transformation mechanisms have been proposed to quench star formation by removing or heating gas in galaxies or their surrounding halos \citep{Dekel+86,DiMatteo+05,Dekel+06,Hopkins+06, Martig+09,Peng+10}. PSBs are ideal laboratories to test some of these mechanisms. Surprisingly, previous studies discovered large amounts of atomic or molecular gas and dust in PSBs \citep{Chang+01,Buyle+06,Zwaan+13, French+15, Rowlands+15, Alatalo+16b,Suess+17, Yesuf+17,Smercina+18,Li+19}. Despite their small sample sizes, these studies reveal a new picture in which star formation is rapidly truncated, but the cold interstellar medium of some of these galaxies is not completely removed or destroyed. The common paradigm, based on simulations of gas-rich major mergers, has been that gas loses angular momentum and is funneled to galaxy centers during such catastrophic events. Intense nuclear starbursts and obscured active galactic nuclei (AGNs) are then triggered. The starburst rapidly depletes most of the cold gas. At the end, the remnant gas and dust are blown out of the host galaxy owing to feedback from the AGN \citep[e.g.,][]{Barnes+91,DiMatteo+05, Hopkins+06, Hopkins+08}. However, in some simulations the effect of AGN feedback in PSBs is inconsequential \citep{Wild+09, Snyder+11, Davis+19}.

\citet{French+15} studied 32 quenched PSBs (QPSBs). Almost all of these QPSBs show signatures of low-ionization nuclear emission-line regions \citep[LINERs; for a review, see][]{Ho08}, and about half of them have CO detections. Those detected in CO have molecular gas content comparable to those of star-forming galaxies (SFGs), while the nondetected PSBs have gas masses more similar to those of quiescent galaxies (QGs). Likewise, \citet{Yesuf+17} studied the molecular gas (CO) properties of 116 PSBs with a variety of AGN and star formation properties. They found that the distribution of molecular gas fractions ($f_\mathrm{H_2} \equiv M_\mathrm{H_2}/M_\star$, where $M_\star$ is the stellar mass) in PSBs is broad and significantly different from that of normal SFGs in the xCOLD GASS survey \citep{Saintonge+17}. PSBs whose active nuclei are classified as Seyferts have median (25\%, 75\%) $f_\mathrm{H_2} =  2.3\,(1.7, 13)\%$, SFGs\footnote{SFGs are defined as galaxies with $\mathrm{\log\,(SSFR/yr}) > -11$ and $\log\,(M_\star/M_\odot) > 10$, where SSFR $\equiv$ SFR/$M_\star$ and SFR is the star formation rate. The $0.15-0.85$ quantile range of $f_\mathrm{H_2}$ is $3\%-14\%$.} have median (25\%, 75\%) $f_\mathrm{H_2} = 7\,(4, 11)\%$, and QPSBs have median (25\%, 75\%) $f_\mathrm{H_2} =  3\, (-, 8)\%$ \citep{Yesuf+17}. The molecular gas fractions generally decrease as the PSBs age. AGN feedback does not seem to affect the cold gas content, at least in PSBs that are in early stages.

\citet{Davis+19} studied the cold interstellar medium of a large sample of PSB galaxies selected from the EAGLE cosmological simulations \citep{Schaye+15}. The formation mechanisms of the simulated PSBs are quite diverse, as are the mechanisms that rapidly deplete gas in these galaxies. Moreover, the simulated PSBs have $10^8-10^{10}\,M_\odot$ of cold star-forming gas (the median mass is $\sim 2 \times 10^{9}\,M_\odot$ at $z < 0.25$). The mechanisms that impact the cold gas and cause the simulated galaxies to become PSBs rapidly include both major and minor mergers, environmental effects, and bursts of AGN activity. 

Directly measuring gas masses for large samples requires huge investments of telescope time. Alternatively, dust absorption or emission can be used to indirectly probe the cold interstellar medium of large samples of galaxies. Rest-frame far-infrared emission measurements, while more widely available than CO observations, are still not easily accessible for large galaxy samples. In \citet{Yesuf+19}, we showed that dust absorption is a dirt-cheap alternative to estimate molecular gas masses for large samples. This method can be applied to a variety of problems (e.g., \citealt{YesufHo20, ZhuangHo20}), of which this current study is but one example. In this work, we analyze the H$\alpha$/H$\beta$ ratios of QPSBs and estimate their molecular gas content using our method. Although this method is indirect and accurate to within a factor of $\sim 2-3$, it enables a complementary, statistical study of molecular gas in a sample of QPSBs that is more than 10 times larger and more homogeneously selected than that of \citet{French+15}. 

In Section~\ref{sec:data}, we describe the data and methodology used to estimate molecular gas. In Section~\ref{sec:res}, we present the gas masses and gas fractions of QPSBs. Section~\ref{sec:disc} discusses our results in a general context of previous studies. A summary of this work and its main conclusions are given in Section~\ref{sec:conc}.

\section{DATA AND METHODOLOGY}\label{sec:data}

This section describes the data used and the sample selection, the spectral stacking method and its validity, our method of estimating molecular gas masses using H$\alpha$/H$\beta$ ratios, and the validity of applying it to study PSBs.

\subsection{SDSS Data and Sample Selection}

Our galaxy sample is taken from the Sloan Digital Sky Survey \citep[SDSS;][]{Abazajian+09,Alam+15}. The publicly available Catalog Archive Server (CAS)\footnote{http://skyserver.sdss.org/casjobs/ \\ We use CAS in the context of data release 15 (DR15) to retrieve various measurements in different catalogs.  However, the sample of galaxies we use is restricted to those in DR7, because the axis ratio measurements from \citet{Simard+11} are only available for DR7 galaxies. The following tables are queried: {\tt photoobjall, galSpecIndx, galSpecInfo, galSpecLine, galSpecExtra, wISE\_xmatch, wISE\_allsky}, and {\tt specDR7.}} is used to compile some of the measurements used in this work, including emission-line fluxes, spectral indices, median stellar masses, and the Petrosian half-light radi. To exclude edge-on galaxies, we additionally use ellipticities/axial ratios derived from single-component S\'{e}rsic fits from the catalog of \citet{Simard+11}. The orientation effect may complicate the estimation of molecular gas mass from dust absorption in edge-on galaxies. 

The SDSS CAS measurements are estimated using the same methods developed by the MPA-JHU group\footnote{https://www.sdss.org/dr15/spectro/galaxy\_mpajhu/}\citep[details can be found in][]{Kauffmann+03a, Brinchmann+04}. Briefly, stellar masses are calculated using a Bayesian methodology and {\it ugriz} photometry alone, but corrected for contamination by nebular emission lines. The emission lines are measured after subtracting the best-fitting stellar population model of the continuum. The emission lines are fit simultaneously as Gaussians. The Balmer lines are required to have the same line width and velocity offset, and so, too, the forbidden lines.

Throughout this work (with the exception of Section~\ref{sec:CO}), we select galaxies that have axial ratios $b/a > 0.5$, redshifts $z = 0.02 - 0.15$, and stellar masses $M_\star = 10^{10}-10^{11}\,M_\odot$. We also require that the median signal-to-noise ratio (S/N) per pixel of the entire spectrum of the galaxy be greater than 10; it is difficult to accurately model the continuum and measure absorption lines when the S/N is smaller than 10. The QPSBs are defined as galaxies with the equivalent width (EW) of H$\alpha < 3$ \,{\AA} (in emission) and H$\delta_A > 4$\,{\AA} (in absorption). Only about 10\% of QPSBs are excluded by the S/N requirement. We visually inspect the optical spectra of the QPSB sample resulting from the above definition and remove several tens of SFG contaminants with bad measurements\footnote{These objects have gaps around their H$\alpha$ continua. Except two objects (with DR7 ID 587726101488795878 and 588011123581124735), the contaminants can be removed automatically by requiring that the EW of H$\beta> 3$\,{\AA} and that there are nonzero H$\alpha$ continuum flux measurements, which are available in the {\tt galSpecLine} catalog.}. For comparison, we define QGs as galaxies with 4000\,{\AA} break $D_n(4000) > 1.6$, H$\alpha < 3$\,{\AA}, and H$\delta_A < 2$\,{\AA} \citep[for the definitions of the indices, see][]{Worthey+97,Balogh+99,Kauffmann+03a}.

Selecting a complete and unbiased sample of the progenitors of QPSBs is not easy. Some attempts have been made to improve the definition of PSBs to include galaxies with ongoing star formation and/or AGN activity \citep{Wild+10,Yesuf+14}. As a comparison sample, here we simply study the inferred gas content of Seyferts with H$\delta_A > 4$\,{\AA} that possess early-type morphologies. We identify Seyferts (in contrast to SFGs) using the emission-line ratios [\ion{O}{3}]/H$\beta$ and [\ion{S}{2}]/H$\alpha$ \citep{Kewley+06}, and we impose the additional cut H$\alpha > 3$\,{\AA}. We use $r$-band concentration index, $C = R_{90}/R_{50} > 2.6$, and stellar mass surface density in $z$ band, $\Sigma_\star = 0.5\,M_\star/(\pi R_{50,z}^2) > 10^{8.5}\,M_\odot$\,kpc$^{-2}$, as indicators of early-type morphology \citep{Strateva+01,Kauffmann+03b}. About 80\% of QPSBs and $\sim 73$ QGs satisfy these morphology cuts, while only $\sim 24$\% of SFGs satisfy them.

\begin{figure}
\includegraphics[width=0.98\linewidth]{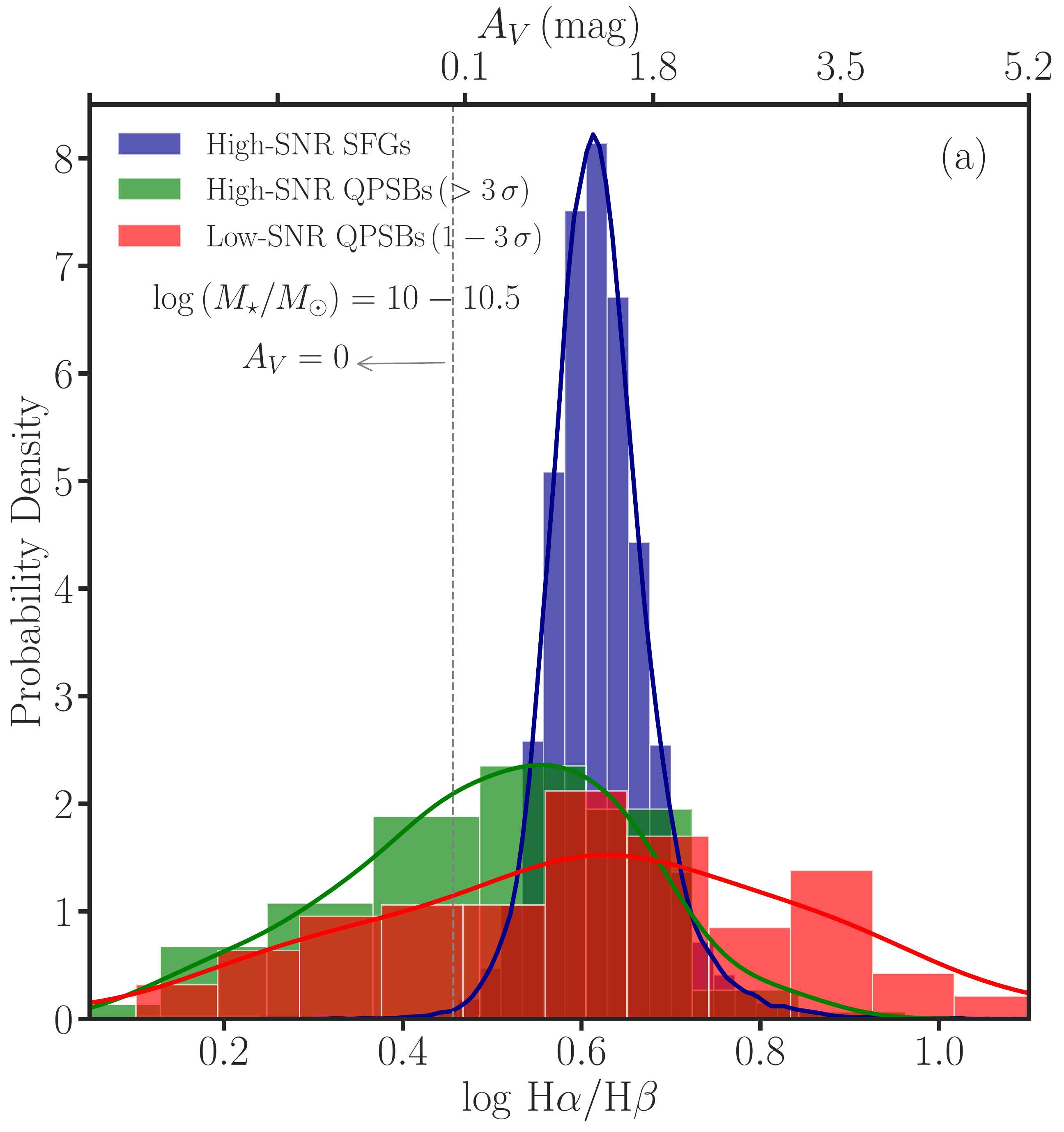}
\includegraphics[width=0.98\linewidth]{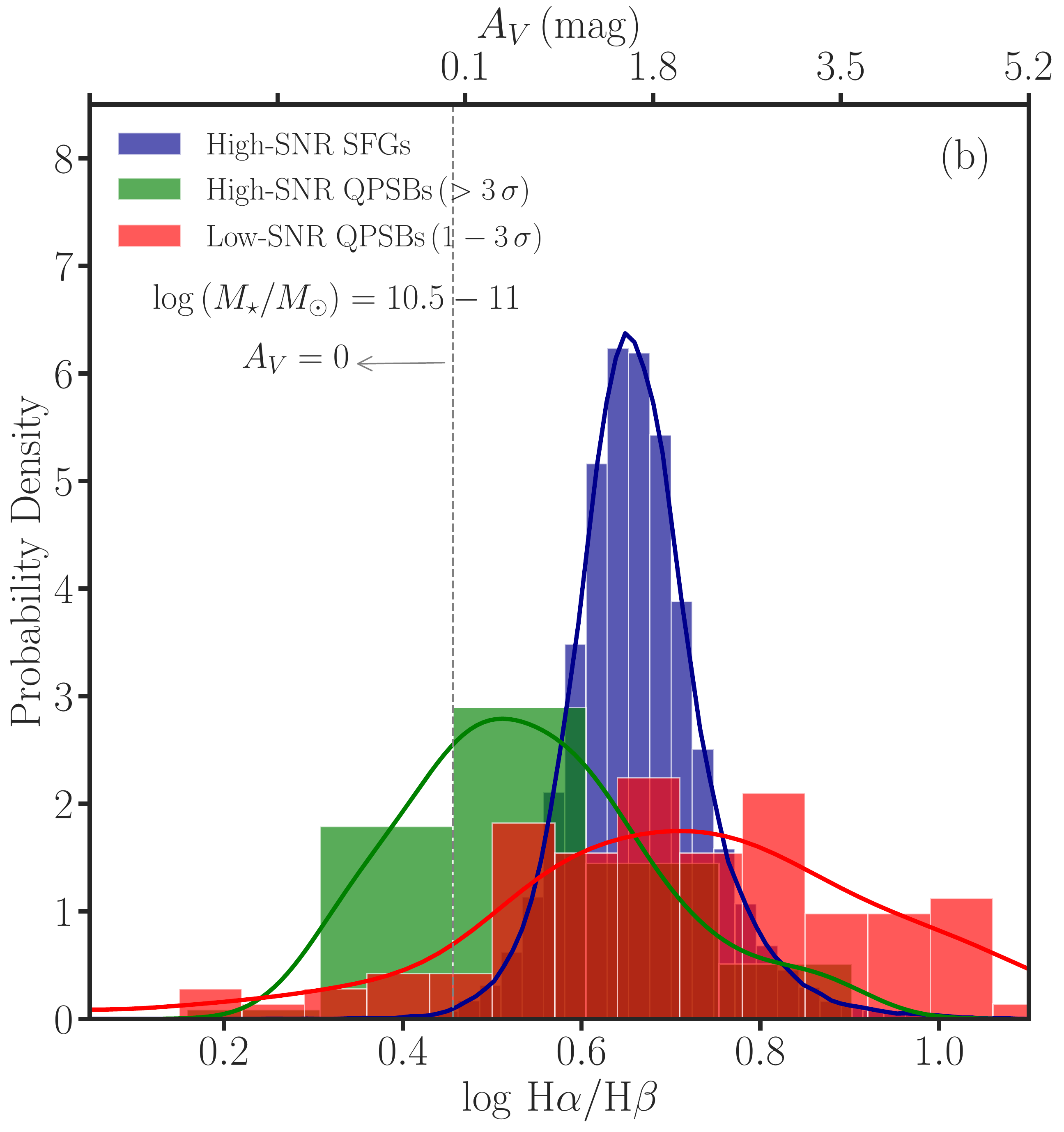}
\caption{Distribution of H$\alpha$/H$\beta$ ratios of face-on ($b/a > 0.5$) QPSBs and the comparison samples of SFGs in two stellar mass ranges. The three curves are kernel density estimates. QPSBs have a wide range of H$\alpha$/H$\beta$ ratios. Some are dust-free while others are as dusty as SFGs. QPSBs whose H$\alpha$/H$\beta$ ratios are not well measured ($< 3\,\sigma$) will be stacked. The H$\alpha$/H$\beta$ ratios measured from the stacked spectra will show that the QPSBs with low- H$\alpha$/H$\beta$ are on average as dusty as SFGs.The high-S/N H$\alpha$/H$\beta$ measurements below the dashed line are underestimated, perhaps due to spectra with insufficient S/N. We set $A_V = 0$\,mag for the objects below this line. \label{fig:hahb_dist}}
\end{figure}

\begin{figure*}
\includegraphics[width=0.98\linewidth]{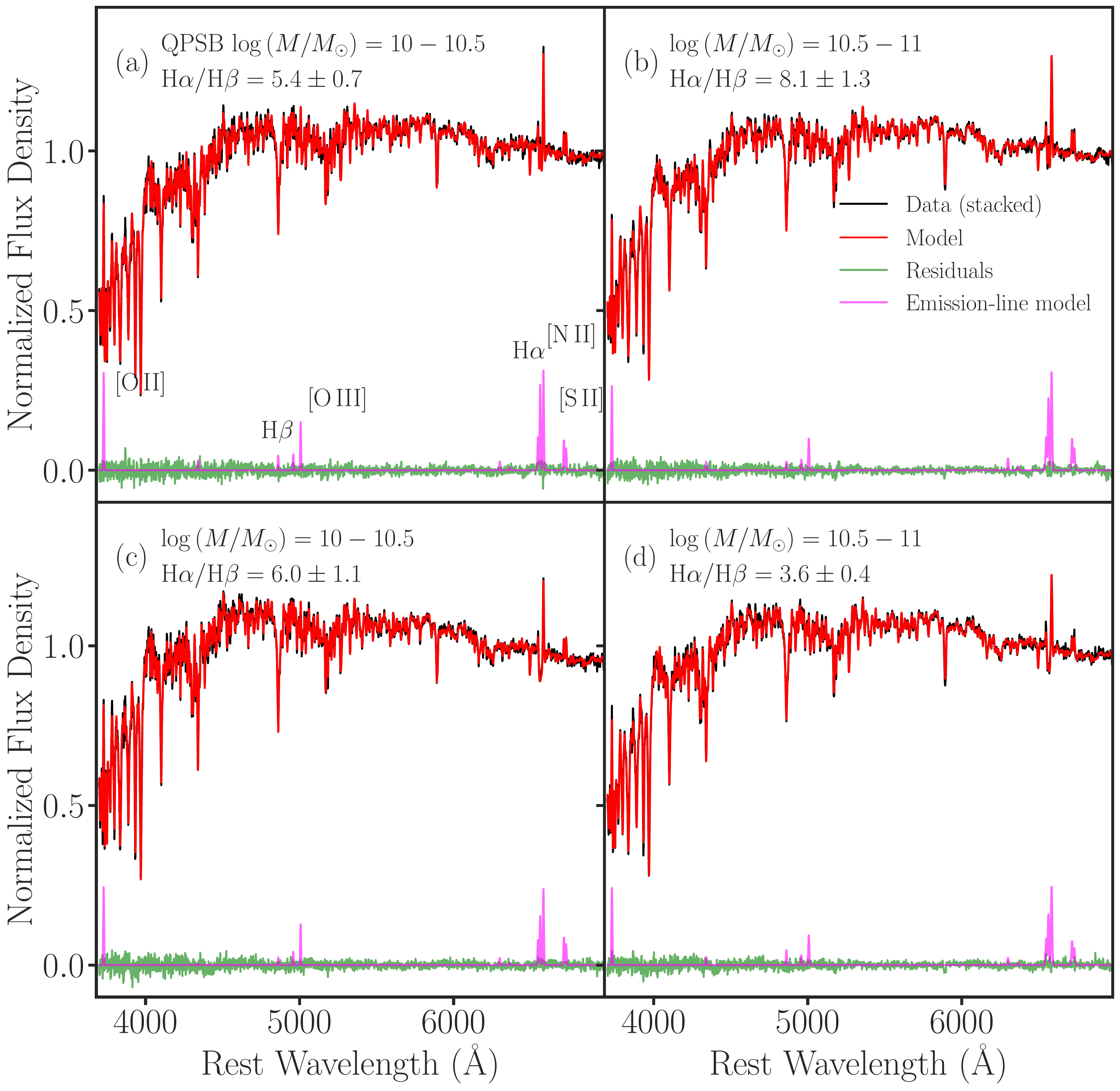}
\caption{Example stacked spectra (black) of subsamples of QPSBs with low-S/N ($1-3\,\sigma$) $\mathrm{H\alpha/H\beta}$ ratios. The model fits for the stellar population plus emission lines are shown in red, and the emission-line model components are shown in magenta. The models reproduce the data well. A significant number of the low-S/N QPSBs have high amounts of dust absorption.\label{fig:coaddspec}}
\end{figure*}

\subsection{Spectral Stacking Analysis}\label{sec:stacking}

Figure~\ref{fig:hahb_dist} shows the distributions of H$\alpha$/H$\beta$ ratios of QPSBs that have well-measured ($>3\,\sigma$) H$\alpha$ and H$\beta$ fluxes and those that have marginal ($1-3\,\sigma$) H$\beta$ detections. About 93\% of all face-on QPSBs (total $N = 535$) have well-measured H$\alpha$ fluxes, but only $\sim 40$\% of the QPSBs have well-measured H$\beta$ fluxes. This fraction increases to $\sim 60$\% or $\sim 80$\% if H$\beta$ detection significance is $>2\,\sigma$ or $>1\,\sigma$. Although we make S/N cuts on the emission-line flux ($>3\,\sigma$) and the continuum measurements (median S/N pixel$^{-1} \, > 10$), the individual measurements of H$\alpha$/H$\beta$ should still be interpreted with caution. Some QPSBs ($\sim 20\%/30$\%) have H$\alpha$/H$\beta < 2.5/2.8$, which are unphysical, and are likely underestimated due to insufficient S/N of the QPSB spectra (S/N pixel$^{-1} \, < 25$) or imperfect deblending of the Balmer emission and absorption lines \citep[see, e.g.,][]{Ho+97}. These measurements merely indicate that these objects do not have detectable dust absorption. Furthermore, Figure~\ref{fig:hahb_dist} suggests that QPSBs with marginally detected H$\beta$ are likely more dusty than those with well measured H$\alpha$ and H$\beta$ fluxes. 

Given the S/N limitation of the spectroscopic observations, we attempt to characterize the distributions of H$\alpha$/H$\beta$, $M_\mathrm{H_2}$, and $f_\mathrm{H_2}$ for the QPSB sample by stacking spectra whose H$\alpha$/H$\beta$ cannot be accurately measured individually. We also demonstrate that this method gives consistent results with the high-S/N H$\alpha$/H$\beta$ measurements based on individual spectra.

To stack the SDSS spectra, each spectrum is shifted to its rest-frame wavelength and is normalized by the mean flux in the wavelength range 5445\,{\AA}--5550\,{\AA}. This range is free of strong emission or absorption lines. The normalization ensures equal weighting of both faint and bright galaxies in the sample. The rest-frame spectra are then interpolated on a logarithmic wavelength grid with a bin of $\Delta \log\,\lambda = 10^{-4}$ dex, which is the same as the pixel spacing of the observed wavelength of SDSS spectra. A median composite spectrum is constructed from the median flux densities at a given wavelength bin of the rest-frame spectra. All spectra are given equal weights. The errors of the median composite spectrum are estimated using bootstrap resampling. Namely, we repeat the above stacking procedure to produce 200 median composite spectra by resampling with replacement using a subset of individual spectra of the same size as the original sample. The standard deviation of the median fluxes of the 200 composite spectra in a given wavelength bin is the error of the median flux at that given bin.

We use the penalized pixel-fitting ({\tt pPXF}) code \citep{Cappellari+04, Cappellari17} to fit the stacked spectra with stellar population and ionized gas emission-line models. After masking out major emission-line regions, the code first determines the optimal linear combinations of simple stellar population templates. The templates are based on the MILES stellar library \citep{Vazdekis+10}, and they have spectral resolutions comparable to those of the SDSS spectra. The {\tt pPXF} code allows the use of additive or multiplicative Legendre polynomials in order to adjust possible continuum mismatch between model templates and observed data. To that end, we use a 10th-order multiplicative polynomial. We have verified that the exact choice of polynomial order does not significantly affect the H$\alpha$/H$\beta$ measurements. Finally, the Balmer and forbidden emission lines are modeled using single-Gaussian line profiles\footnote{In detail, some profiles are more complicated than a single Gaussian, but for our purposes this assumption suffices.}. All Balmer lines have a same velocity dispersion, and so, too, the forbidden emission lines, but not necessarily the same velocity dispersion as that of the Balmer lines. Prior to the least-squares minimization, the model emission lines are convolved with the SDSS spectral resolution. Note that the flux errors calculated from the bootstrap scheme are also used as inputs to {\tt pPXF}. Example co-added spectra of low-S/N QPSBs and their fits are shown in Figure~\ref{fig:coaddspec}.

Table~\ref{tbl:stackHSN} shows that the distributions of H$\alpha$/H$\beta$ resulting from the stacking analysis and individual spectra are generally consistent. The stacking analysis indicates that H$\alpha$/H$\beta$ ranges between $\sim 3-5$ for the high-S/N QPSB sample. Based on the stacked analysis and the expectation from case B recombination physics, we deem the individual measurements of H$\alpha$/H$\beta$ below $\sim 2.9$ inaccurate (underestimated). We use $A_V = 0$\,mag for the high-S/N QPSBs with H$\alpha$/H$\beta < 2.9$ when estimating their molecular gas masses. Our conclusions are robust despite some individual QPSBs having inaccurate H$\alpha$/H$\beta$ measurements. This is because how exactly close their H$\alpha$/H$\beta$ ratios are to the dust-free, intrinsic ratio ($\sim 2.9-3.1$) do not cause significantly different gas mass estimates, and we also confirm the gas estimates using the stacking analysis.


\begin{deluxetable*}{lccccccc}
\tabletypesize{\footnotesize}
\tablecolumns{8} 
\tablewidth{0pt}
\tablecaption{Dust and Molecular Gas in High-S/N QPSBs \label{tbl:stackHSN}}
\tablehead{
\colhead{ID} & \colhead{$\log\,(M_\star/M_\odot$)} &  \colhead{H$\alpha$/H$\beta$}  & \colhead{Stack H$\alpha$/H$\beta$} & \colhead{$\log\,(M_\mathrm{H_2}/M_\odot)$} &  \colhead{Stack $\log\,(M_\mathrm{H_2}/M_\odot)$} & \colhead{$\log\,f_\mathrm{H_2}$} & \colhead{$N$ Galaxies} \\
\colhead{(1)} &
\colhead{(2)} &
\colhead{(3)} &
\colhead{(4)} &
\colhead{(5)} &
\colhead{(6)} &
\colhead{(7)} &
\colhead{(8)} 
}
\startdata
1 & $10.5 -11$ & $3.3\,(2.6, 4.7)$ & $4.1 \pm 0.3$ & $8.9\,(8.7, 9.3)$  &  $9.2\,(9.0, 9.3)$ & $-1.8\,(-2.0, -1.4)$ &"79 \\ 
1a & $10.5 -11$ & $4.7\,(4.1, 6.4)$ & $4.6 \pm 0.5$ & $9.2\,(9.1, 9.6)$ & $9.3\,(9.1, 9.5)$ & $-1.4\,(-1.7, -1.1)$ &"27 \\   
1b & $10.5 -11$ & $3.3\,(3.1, 3.7)$ & $4.0 \pm 0.5$ & $8.9\,(8.8, 9.1)$ & $9.1\,(8.9, 9.3)$ & $-1.8\,(-1.9, -1.6)$ &"26 \\  
1c & $10.5 -11$ & $2.6\,(2.2, 2.9)$ & $3.2 \pm 0.4$ & $8.8\,(8.7, 8.9)$ & $8.9\,(8.7, 9.1)$ & $-1.9\,(-2.1, -1.8)$ &"26 \\  
\hline
2  & $10 -10.5$ & $3.2\,(2.1, 4.5)$ & $3.7 \pm 0.2$ & $8.7\,(8.5, 9.0)$ & $8.8\,(8.7, 9.0)$ & $-1.6\,(-1.8, -1.2)$ & 125 \\  
2a & $10 -10.5$ & $4.6\,(4.1, 5.2)$ & $5.3 \pm 0.6$ & $9.0\,(8.9, 9.3)$ & $9.3\,(9.1, 9.4)$ & $-1.2\,(-1.4, -1.0)$ &"35 \\
2b & $10 -10.5$ & $3.5\,(3.2, 3.7)$ & $3.3 \pm 0.3$ & $8.7\,(8.6, 8.9)$ & $8.7\,(8.5, 8.8)$ & $-1.5\,(-1.7, -1.4)$ &"35 \\ 
2c & $10 -10.5$ & $2.3\,(1.7, 2.8)$ & $2.7 \pm 0.3$ & $8.6\,(8.5, 8.7)$ & $8.6\,(8.5, 8.7)$ & $-1.7\,(-1.8, -1.6)$ &"55 \\ 
\enddata 
\tablecomments{The two samples without letter identification include all galaxies in their corresponding stellar mass ranges. Subsamples with label ``c'' have H$\alpha$/H$\beta < 3$ (i.e., low dust). Subsamples with label ``a'' and ``b'' split galaxies with H$\alpha$/H$\beta > 3$ into two equal bins. The thresholds to split galaxies to medium and high dust content are 3.9 and 4 for the high-mass and low-mass ranges, respectively. We use the notation $X\,(Y, Z)$ to denote $X =$ median (50\%), $Y = 15\%$, and $Z= 85\%$ of a distribution. Columns (3), (5), and (7) give results that are based on individual spectra. The gas estimates use the median scaling relation. In comparison, Columns (4) and (6) present results that are based on stacked spectra. The two approaches give similar results.}
\end{deluxetable*}

\subsection{Estimation of Gas Masses from SDSS Data}\label{ssec:gas}

We calculate the $V$-band dust absorption using the observed H$\alpha$/H$\beta$ ratio and the dust attenuation curve of \citet[][see also \citet{Wild+11a}]{Charlot+00} as follows:

\begin{equation}
Q_\lambda = 0.6\,(\lambda/5500)^{-1.3} + 0.4\,(\lambda/5500)^{-0.7}.
\end{equation}

\noindent
Assuming that the intrinsic Balmer decrement is H$\alpha$/H$\beta$ = 2.86 for inactive galaxies and H$\alpha$/H$\beta$ = 3.1 for AGNs \citep[e.g.,][]{Ferland+83,Gaskell+84},

\begin{equation}\label{eq:AV}
 A_{V} = \frac{2.5}{(Q_{4861} - Q_{6563})} \times \log \frac{\mathrm{H}\alpha/\mathrm{H}\beta}{3.1 \,\mathrm{or} \, 2.86},
\end{equation}

\noindent where $Q_{4861} - Q_{6563} = 0.31$. If the observed ratio of an object is below the intrinsic ratio (2.86/3.1), we set $A_V = 0$ mag\footnote{This might cause the object's predicted gas mass to be lower by up to $\sim 0.1$ dex than it actually is. Such small changes are not important for our application. An object with H$\alpha$/H$\beta < 2.9-3.1$ might have $\log\, (M_\mathrm{H_2}/M_\odot) \lesssim 8.5-8.8$. For our purpose, knowing that the object has a gas mass similar to QGs is sufficient. We also use 2.86 as the intrinsic ratio when H$\alpha$/H$\beta$ ratios are measured from stacked spectra.}. For reference, $A_V = [0.5,1,1.5, 2, 3, 3.5]$ mag corresponds to $\mathrm{H\alpha/H\beta} \approx [3.3, 3.8, 4.4, 5.1, 6.8, 7.8]$.

We estimate the molecular gas masses of PSBs using the empirical estimator presented in our previous work \citep{Yesuf+19}. In particular, we use dust attenuation (H$\alpha$/H$\beta$ ratio), the average gas-phase metallicity ($Z$), which is inferred from the stellar mass-metallicity relation \citep{Tremonti+04}, and the Petrosian half-light radius ($R_{50}$) to estimate $M_\mathrm{H_2}$, which would be within $\sim 2.5$ times the true $M_\mathrm{H_2}$ if observed directly. In \citet{Yesuf+19}, we analyzed the molecular (and atomic) gas data of local representative galaxies ($M_\star = 10^{9}-10^{11.5}$ \msun\, and $z = 0.01 - 0.05$) from the extended GALEX Arecibo SDSS Survey \citep[xGASS; ][]{Catinella+18} and xCOLD GASS \citep{Saintonge+17}. We fitted censored quantile regression to summarize the median and the $0.15/0.85$ quantile multivariate relationships among $M_\mathrm{H_2}$, {\AV}, $Z$, and/or $R_{50}$. These empirical relations are also consistent with independent gas data of galaxies in the Herschel Reference Survey compiled by \citet{Boselli+14a}.

For a QPSB galaxy whose $\mathrm{H\alpha/H\beta}$ ratio is measurable with high significance, its own H$\alpha$/H$\beta$ ratio is used to estimate its molecular gas mass. On the other hand, for a galaxy whose H$\alpha$/H$\beta$ ratio is not well measured, we use the H$\alpha$/H$\beta$ value obtained from the stacked spectrum to which the galaxy contributed. Given the average $\langle \mathrm{H\alpha/H\beta} \rangle$ and its error $\sigma_{\mathrm{H\alpha/H\beta}}$ from fitting the stacked spectrum, we create, for each galaxy in the stack, 1000 random samples of H$\alpha$/H$\beta$ from a uniform distribution in the interval [$\langle \mathrm{H\alpha/H\beta} \rangle$ - $\sqrt{3}\,\sigma_\mathrm{H\alpha/H\beta}$, $\langle \mathrm{H\alpha/H\beta} \rangle$ + $\sqrt{3}\,\sigma_\mathrm{H\alpha/H\beta}$]. Then, we calculate 1000 samples of $A_V$ and the corresponding $M_\mathrm{H_2}$ values for each galaxy using its stellar mass and radius, as outlined above. Finally, we combine all $\log\, M_\mathrm{H_2}$ samples ($1000 \times n$, where $n$ is the number of objects in a given stack) and calculate the median (15\%, 85\%) of the combined $\log\, M_\mathrm{H_2}$ as approximate summary statistics of the ``true'' distribution of $\log\, M_\mathrm{H_2}$ for galaxies in the stack. Using several subsamples of galaxies with well-measured H$\alpha$/H$\beta$ ratios, we show that our method of approximating the actual distributions of $\log\, M_\mathrm{H_2}$ using the stacked spectra results in reasonably accurate inference. Table~\ref{tbl:stackHSN} also compares the $M_\mathrm{H_2}$ distributions resulting from the stacking analysis and individual spectra.

We adopt two procedures to further capture and quantify additional galaxy-to-galaxy variations of $\mathrm{H\alpha/H\beta}$ ratios. In the first procedure, we create multiple stacked spectra by subdividing galaxies whose H$\alpha$/H$\beta$ ratios are not well measured into subsamples using the limited information in H$\alpha$/H$\beta$: poorly measured H$\alpha$/H$\beta$ ($<1\,\sigma$), and  marginally measured high, medium and low H$\alpha$/H$\beta$ ($1-3\,\sigma$).  The second procedure uses the ratios of Wide-field Infrared Survey Explorer \citep[WISE\footnote{The cross-matched WISE data are available in SDSS DR15 tables {\tt wISE\_xmatch} and {\tt wISE\_allsky.}}][]{Wright+10} 12\,$\mu$m to 4.6\,$\mu$m flux density, $\fwise$, to subdivide the stacked sample. \citet{Yesuf+17} found that this flux density ratio is tightly correlated with $f_{\rm{H_2}}$ and is a good proxy for the average gas fraction for both PSBs and non-PSBs.

Table~\ref{tbl:stackHSN} shows that the dispersion of the $M_\mathrm{H_2}$ distribution due to galaxy-to-galaxy variations can be teased out when the sample is separately stacked in three H$\alpha$/H$\beta$ ranges. Stacking of all galaxies in the high-S/N QPSB sample nevertheless gives a valid and useful average (statistical) description of the
$M_\mathrm{H_2}$ distribution.

In an attempt to \emph{approximately} reconstruct the $\log\, M_\mathrm{H_2}$ distribution of all QPSBs, we combine the information from the individual measurements of $\log\, M_\mathrm{H_2}$ with those from the stacking analysis, as follows. For each independent $i$ subsample of stacked spectra, we generate $1000 \times n_i$ samples of $\log\, M_\mathrm{H_2}$, as described above using the mean H$\alpha$/H$\beta$ ratios. We then fit the distribution of $\log\, M_\mathrm{H_2}$ for each subsample with a lognormal function. When the analysis of all the subsamples is done, we draw 500 times samples of size $n_i$ at a time from each $i$ lognormal distribution. We then combine all these draws with 500 exact replications of the array of individual $\log\, M_\mathrm{H_2}$ measurements. The final $\log\, M_\mathrm{H_2}$ sample is binned and normalized to give a probability density function. Similarly, to reconstruct the probability density function of $\log\, f_\mathrm{H_2}$ a Gaussian distribution is used instead of a lognormal distribution. When the stacking analysis results in a lower limit on H$\alpha$/H$\beta$, $M_\mathrm{H_2}$ and $f_\mathrm{H_2}$ are sampled uniformly within 0.5 dex above the corresponding gas lower limits. The results do not change qualitatively if instead galaxies with lower limits are sampled from (added to the numbers of) the closest detected subsamples.

\begin{figure*}
\includegraphics[width=0.95\linewidth]{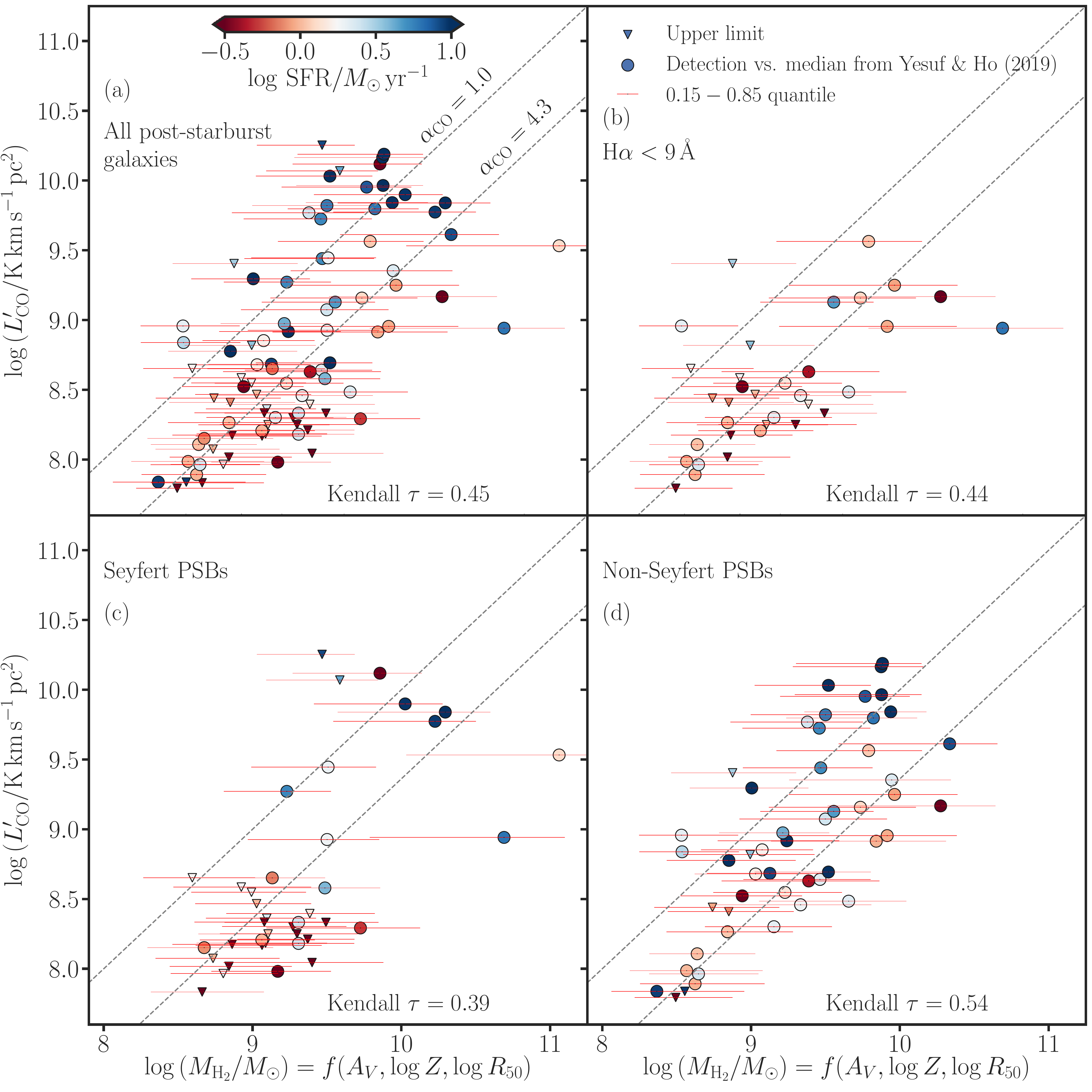}
\caption{Comparison of the observed CO luminosities of $\sim 100$ PSB galaxies \citep{Yesuf+17} with the predicted molecular gas masses ($M_\mathrm{{H_2}}$), using the gas scaling relation \citep{Yesuf+19} with dust absorption ({\AV}), average metallicity ($Z$), and half-light radius ($R_{50}$). The points are color-coded by SFR \citep{Salim+16, Salim+18}. The circles denote galaxies with detections, and the triangles indicate gas mass upper limits. The x-axis positions of the detections and upper limits correspond to the median predictions; the red error bars show the predicted $0.15 - 0.85$ quantile ranges. Panel (a) shows the entire sample of PSBs, and the other panels show various subsets: (b) PSBs with low H$\alpha$ EW, (c) Seyfert PSBs, and (d) non-Seyfert PSBs. The dotted lines mark constant $\alpha_{\rm CO} \equiv M_\mathrm{{H_2}}/L^\prime_\mathrm{CO} =$ 1 or 4.3 $M_\odot\,(\mathrm{K\,km\,s^{-1}\,pc^{2}})^{-1}$. Considering the well-known uncertainty of converting CO luminosity to $M_\mathrm{{H_2}}$, this figure demonstrates that it is reasonable to apply our $A_V$-based dust-to-gas scaling relation to PSB galaxies. Our method works well for QPSBs, provided that their $\alpha_\mathrm{CO} \approx 4.3$ $M_\odot\,(\mathrm{K\,km\,s^{-1}\,pc^{2}})^{-1}$. \label{fig:AV_CO}}
\end{figure*}

\subsection{Validating Gas Mass Estimator Using PSBs with CO Data}\label{sec:CO}

We compile a heterogenous sample of PSB galaxy candidates \citep{French+15, Rowlands+15, Alatalo+16b, Yesuf+17} to show that our method presented in \citet{Yesuf+19} can be applied to PSBs. The sample is heterogenous because the sample selection is different in different works, and the whole sample has a wide range of star formation and AGN properties (Figure 1 in \citealt{Yesuf+17}). About $\sim 100$ ($\sim 85$\%) of the PSBs have well-measured H$\alpha$/H$\beta$ ratios. The sample has the following median (15\%, 85\%) properties: (1) $\log \,(M_\star/M_\odot) =  10.5\,(10.1, 10.9)$,  (2) $z = 0.05\,(0.03, 0.13)$, (3) H$\delta_A = 5.6\,(4, 7)$ \AA, and (4) EW(H$\alpha$) = 9\,(2, 31) \AA. Thus, at least half have either significant amounts of star formation or strong AGN activity. Among those that are classifiable, the Seyfert AGN fraction is $\sim 40\%$. Note that about 15\% of the sample have H$\delta_A < 4$\,{\AA}, but their near-ultraviolet-optical colors indicate that these galaxies are likely aged PSBs \citep{Yesuf+17}. We exclude one galaxy from \cite{French+15} labeled H01, because resolved Atacama Large Millimiter Array (ALMA) CO imaging showed that the gas is associated with a companion, not with the QPSB \citep{French+18}.

Figure~\ref{fig:AV_CO}a compares the $M_\mathrm{H_2}$ predicted by our empirical estimator and the observed CO luminosities of both quenched and quenching sample of PSB galaxies. The first panel (a) shows all galaxies in the sample, while the other panels split the sample into (b) low-SFR galaxies, (c) Seyferts, and (d) non-Seyferts. The figure is color-coded by SFR in order to show that the scatter is not random. The SFR is derived by spectral energy distribution fitting of ultraviolet, optical, and mid-infrared photometry \citep{Salim+16, Salim+18}. Salim et al. deem their SFRs reliable even for PSBs, and that they are more likely to represent true levels of star formation averaged over the last 100 Myr for QPSBs. The SFRs are used to make supplementary points and are not essential for this work. The trend with SFR is also qualitatively similar if we instead use SFRs (upper limits) estimated from the WISE $12\,\mu$m luminosity \citep{Hayward+14, Cluver+17}.

The predicted median $M_{\mathrm{H_2}}$ is moderately correlated with the observed CO luminosity (Kendall $\tau \approx 0.5$; Spearman $\rho \approx 0.6-0.7$)\footnote{The upper limits are used in computing Kendall $\tau$.}. The ratios of the predicted median molecular masses to the observed CO luminosities give a median (15\%, 85\%) CO conversion factor of $\alpha_\mathrm{CO} \approx 4.8 \,(1, 15)\, M_\odot\,(\mathrm{K\,km\,s^{-1}\,pc^{2}})^{-1}$. Overall, the predicted median gas masses are in good agreement with the observed CO luminosities, provided that $\alpha_\mathrm{CO} \approx 4\, M_\odot\,(\mathrm{K\,km\,s^{-1}\,pc^{2}})^{-1}$ for most PSBs.

As shown in Figure~\ref{fig:AV_CO}b, the PSBs with H$\alpha$ EW $< 9$\,{\AA} (half of the sample) lie close to the line of $\alpha_\mathrm{CO} = 4.3\, M_\odot\,(\mathrm{K\,km\,s^{-1}\,pc^{2}})^{-1}$. Therefore, our method works well for QPSBs, without appealing to $\alpha_\mathrm{CO}$ values that are significantly different from that of the Milky Way. Similarly, for $\alpha_\mathrm{CO} \approx 4.3\, M_\odot\,(\mathrm{K\,km\,s^{-1}\,pc^{2}})^{-1}$, the predicted median gas masses for Seyfert PSB candidates are also consistent with the observed CO luminosities (Figure~\ref{fig:AV_CO}c). Likewise, about two-thirds of the PSBs that are not Seyferts (in Figure~\ref{fig:AV_CO}d) are also consistent with having $\alpha_\mathrm{CO} \approx 4.3\, M_\odot\,(\mathrm{K\,km\,s^{-1}\,pc^{2}})^{-1}$. They are also likely low-SFR galaxies. But for the third of the PSBs that are not Seyferts, shown in Figure~\ref{fig:AV_CO}d, either our mean relation underpredicts their gas masses or their $\alpha_\mathrm{CO} \approx 1\, M_\odot\,(\mathrm{K\,km\,s^{-1}\,pc^{2}})^{-1}$. We think that the latter is more likely. If, instead, the former is true, it would not weaken our conclusions; the trend goes in the opposite direction, and some of the PSBs will actually have more gas than predicted. The SFRs of some of the PSBs with CO data indicate that these galaxies lie above the ridge of the star-forming main sequence (i.e., they are starburst-like, and some of them are morphologically very disturbed). Even with $\alpha_\mathrm{CO} = 1\, M_\odot\,(\mathrm{K\,km\,s^{-1}\,pc^{2}})^{-1}$, the molecular gas fractions of most of these galaxies are too high ($\gtrsim 30\%$) to be consistent with nonstarburst galaxies of similar stellar masses \citep{Saintonge+17}. Could they just be misidentified starbursts? Supplementary analysis of the CO sample is given in the Appendix~\ref{app:COPSB}.

In their simulation, \citet{Renaud+19} found that $\alpha_\mathrm{CO}$ changes with time after a starburst (i.e., with evolution of the SFR). In addition to the prediction uncertainty of our method, the possible evolution of $\alpha_\mathrm{CO}$ also contributes to the scatter in Figure~\ref{fig:AV_CO}. In agreement with what is shown in Figure~\ref{fig:AV_CO}, the results of \citet{Renaud+19} indicate that $\alpha_\mathrm{CO} \approx 4-6\,M_\odot\,(\mathrm{K\,km\,s^{-1}\,pc^{2}})^{-1}$ is more appropriate for QPSBs than $\alpha_\mathrm{CO} \approx 1\,M_\odot\,(\mathrm{K\,km\,s^{-1}\,pc^{2}})^{-1}$ observed in starbursts and ultra-luminous infrared galaxies \citep[][and references therein]{Bolatto+13}. 

CO\,(1--0) luminosities are estimated from CO\,(2--1) assuming that their ratio is unity for 24 galaxies \citep{Yesuf+17}. This assumption may be yet another source of uncertainty in Figure~\ref{fig:AV_CO}.

To summarize: given the large uncertainty of $\alpha_\mathrm{CO}$ and the heterogeneity of the PSB sample in previous studies, we do not find a compelling indication that our method is inconsistent with these observations. In fact, the success of our gas estimator in reasonably recovering the existing data for the majority of PSBs gives us confidence to apply it in this work.

\section{RESULTS}\label{sec:res}

Table~\ref{tbl:fgas} presents the summary statistics of the distributions of H$\alpha$/H$\beta$, $M_\mathrm{H_2}$, and $f_\mathrm{H_2}$ for the face-on QPSB sample ($N=204$) with well-measured ($>3\,\sigma$) H$\alpha$ and H$\beta$ emission lines. For comparison, the table also presents the summary statistics of SFGs, early-type Seyferts ($C > 2.6$, and $\Sigma_\star > 10^{8.5}\,M_\odot$\,kpc$^{-2}$), and QGs. Because the median $f_\mathrm{H_2}$ depends on stellar mass \citep{Saintonge+17}, we split the sample into two mass bins. As we will show soon, this high-S/N sample is likely biased against more dusty QPSBs. Nevertheless, it is already clear that QPSBs have a wide range of dust absorption (H$\alpha$/H$\beta \approx 3-5$ or $A_V \approx 0-1.5$ mag for the high-S/N sample), and, by inference, molecular gas content; most are gas-poor, while some are remarkably dusty and gas-rich. For example, for the mass range $\log \,(M_\star/M_\odot) = 10.5 - 11$, the observed H$\alpha$/H$\beta$ distributions, using our gas scaling relation described in Section~\ref{ssec:gas}, correspond to median (15\%, 85\%) $\log\,(M_\mathrm{H_2}/M_\odot)$ of $8.9\,(8.7, 9.3)$ and median (15\%, 85\%) $\log\,f_\mathrm{H_2}$ of $-1.8\,(-2.0, -1.4)$. Furthermore, there are significant overlaps among the $f_\mathrm{H_2}$ distributions of QPSBs, SFGs, Seyferts, and QGs for the two mass ranges. However, in general, the medians of the distributions progressively shift toward lower $f_\mathrm{H_2}$, from SFGs ($\sim 5\%-8\%$) to Seyferts ($\sim 4\%-5\%$) to QPSBs ($\sim 2$\%) to QGs ($\sim 1\%-2\%$).

\begin{figure*}
\centering
\includegraphics[width=0.98\linewidth]{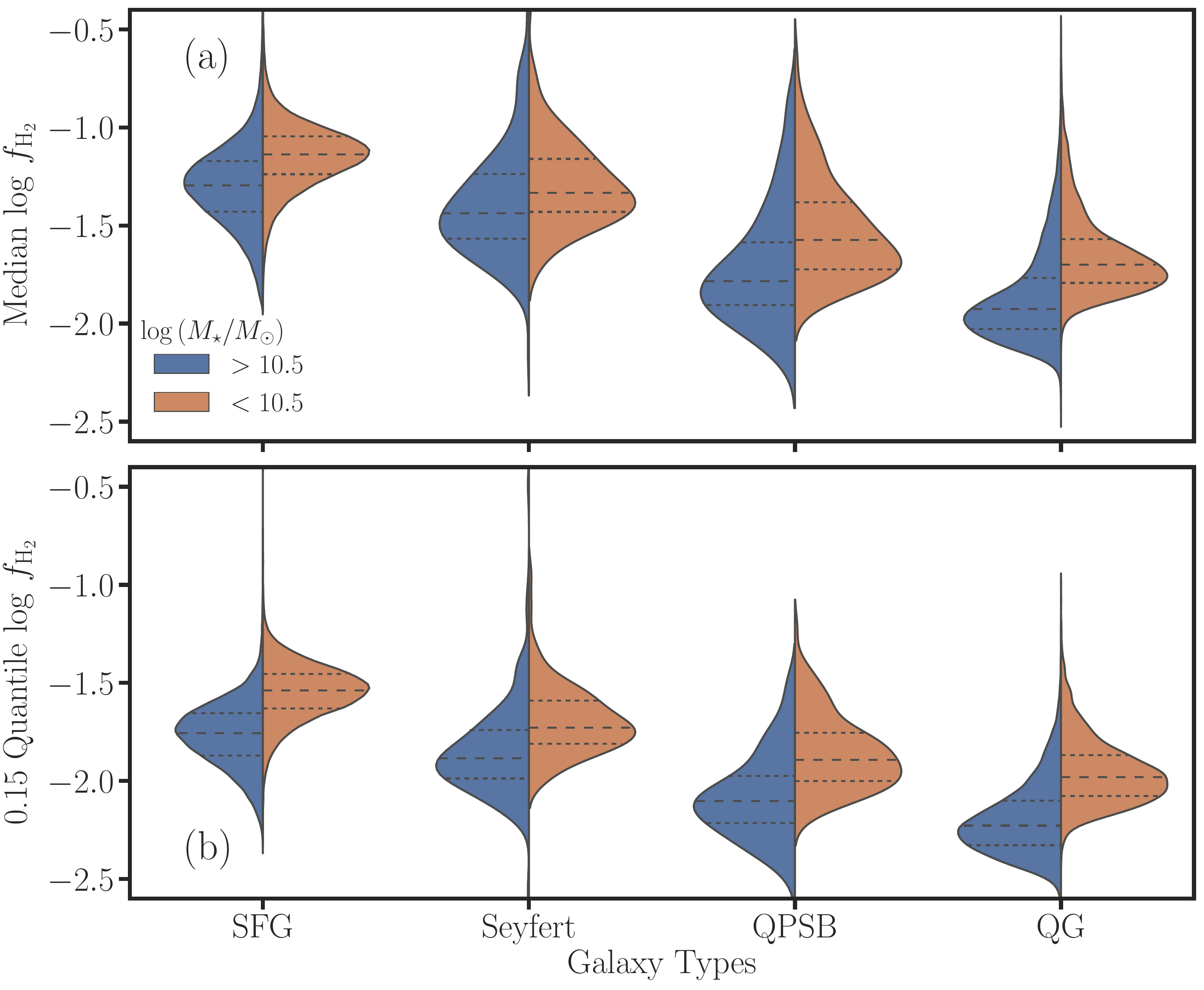}
\caption{Comparison of the distributions of (a) the median and (b) 0.15 quantile of the molecular gas fractions ($f_\mathrm{H_2}$) for SFGs, early-type Seyferts, and QGs. The distributions are compared in two mass bins. All galaxies have $\log \,(M_\star/M_\odot) = 10 - 11$ and are required to have well-measured H$\alpha$ and H$\beta$ emission lines. The dashed or dotted lines on each distribution denote the median or the interquartile ($25\%-75\%$) range. The PSBs have a wide range of $f_\mathrm{H_2}$, but the medians/modes of their $f_\mathrm{H_2}$ distributions are intermediate between those of SFGs and QGs. Later, we will improve the QPSB distribution by stacking QPSBs with low-S/N H$\alpha$/H$\beta$ ratios, which are missing from the upper half of the $f_\mathrm{H_2}$ distribution. \label{fig:VP}}
\end{figure*}

Furthermore, Figure~\ref{fig:VP} shows violin plots of the median and 0.15 quantile $f_\mathrm{H_2}$ distributions of QPSBs in comparison with those of SFGs, early-type Seyferts, and QGs. The distributions are visualized separately in the two mass ranges, $\log \,(M_\star/M_\odot) = 10.5 - 11$ (blue) and $\log \,(M_\star/M_\odot) = 10 - 10.5$ (orange). All galaxies plotted here have individually well-measured H$\alpha$ and H$\beta$ emission lines.

Having shown in Table~\ref{tbl:stackHSN} that the stacking analysis can be confidently applied to study objects with low-S/N ($< 3\,\sigma$) H$\alpha$/H$\beta$ ratios, Table~\ref{tbl:stackLSN} presents the summary statistics of the distributions of H$\alpha$/H$\beta$, $M_\mathrm{H_2}$, and $f_\mathrm{H_2}$ for the low-S/N QPSB sample ($N=327$). In addition to stacking all galaxies with low-S/N H$\alpha$/H$\beta$ ratios, we further subdivide the sample into three or four bins. In most cases, the stacking analysis reveals subsamples that have significantly enhanced dust and molecular gas content compared to  the high-S/N QPSB sample. For example, for the stellar mass range $\log \,(M_\star/M_\odot) = 10.5 - 11$, the mean H$\alpha$/H$\beta \approx 8$\,($A_V \approx 3.6$\,mag). About a third to a half of the QPSBs ($N \gtrsim 50-80$) in this stacked sample might have $M_\mathrm{{H_2}} \gtrsim 10^{10}\,M_\odot$ and $f_\mathrm{H_2} \gtrsim 10\%-20\%$.

To further quantify the dispersion in gas content in the stacked galaxies, we repeat the stacking analysis in Table~\ref{tbl:stackWL} by subdividing the sample using the WISE $\fwise$ ratios, and then measuring H$\alpha$/H$\beta$ from the stacked spectra (see also Appendix~\ref{app:WISE}). This analysis also indicates that most of the low-S/N QPSB subsamples (3/4) have average H$\alpha$/H$\beta \approx 5-8$ and average $f_\mathrm{H_2} \approx 5\%-20$\%.

\begin{figure*}
\includegraphics[width=0.48\linewidth]{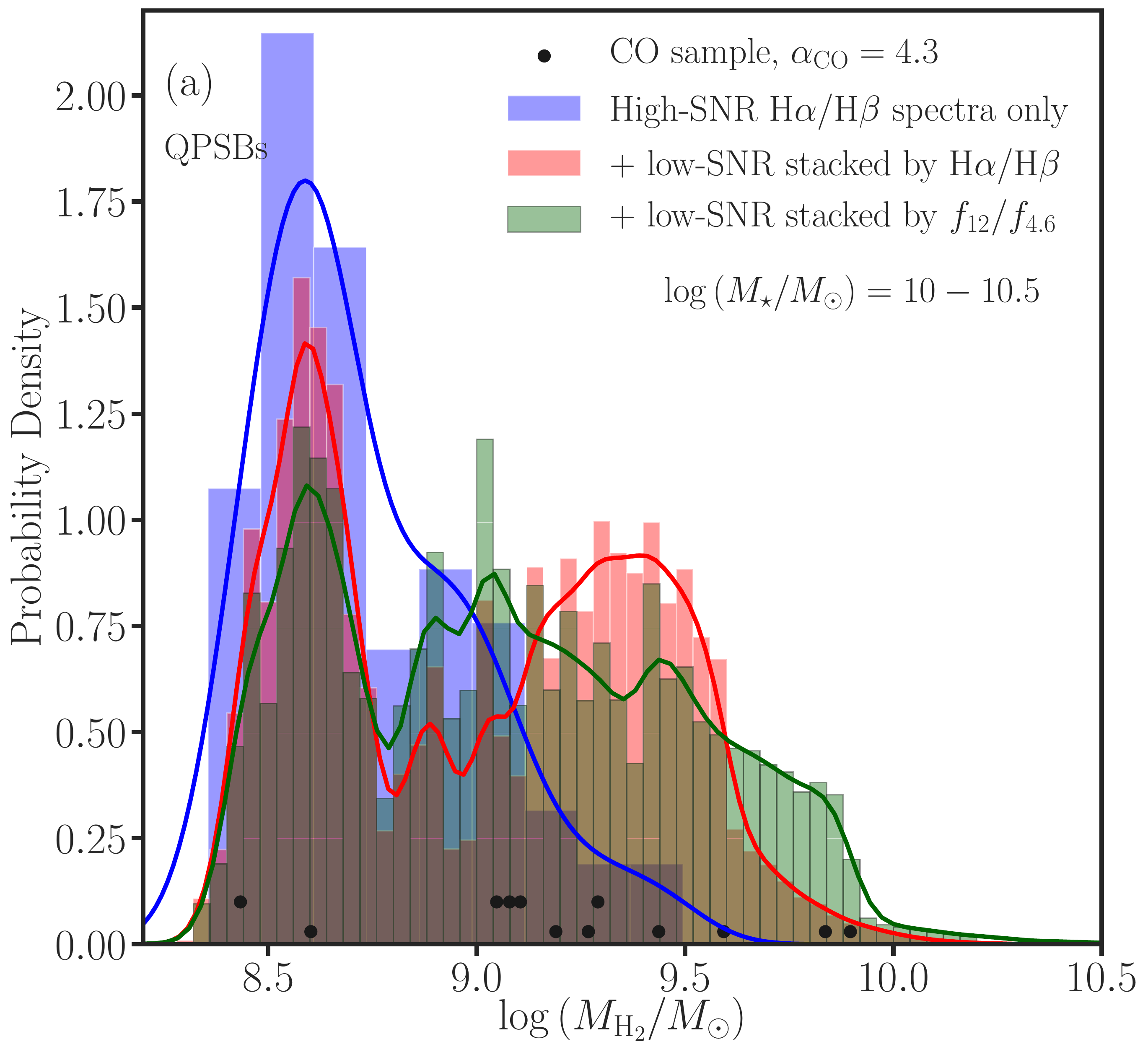}
\includegraphics[width=0.48\linewidth]{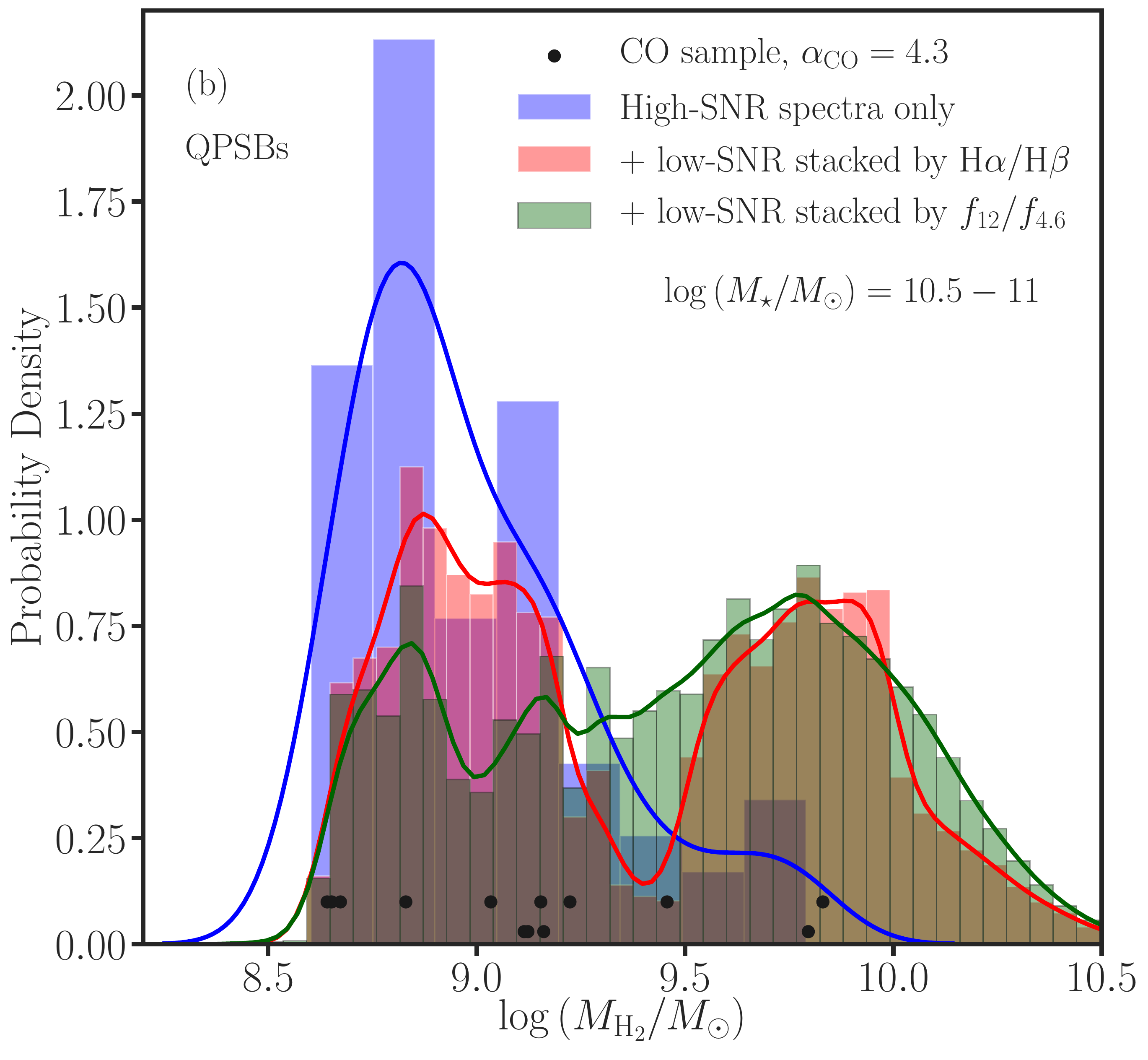}
\caption{Attempted reconstruction of molecular gas distribution of QPSBs at $z=0.02-0.15$. Panels (a) and (b) represent different stellar mass ranges. The figure combines the inferred gas masses of QPSBs with well-measured H$\alpha$/H$\beta$ ratios and of those whose H$\alpha$/H$\beta$ ratios can only be estimated from stacked spectra. Before stacking, the latter QPSBs are subdivided into multiple subsamples by the WISE $\fwise$ ratios or by the partial information in their low-S/N H$\alpha$/H$\beta$ ratios (see the main text for details). The green and red histograms show the $M_\mathrm{H_2}$ probability density functions of the combined samples of individual (blue) and stacked QPSBs. The three curves overlaid on the histograms are kernel density estimates. The black points denote the $M_\mathrm{H_2}$ values of the few QPSBs with CO data. The top points are CO upper limits, while the bottom ones are CO detections. In agreement with CO measurements, the Balmer decrements of QPSBs indicate that the gas content of QPSBs is diverse.  A substantial fraction of QPSBs are likely as gas-rich as SFGs.\label{fig:Mh2_stack}}
\end{figure*}

Figure~\ref{fig:Mh2_stack} shows the reconstructed $M_\mathrm{H_2}$ distributions of QPSBs with the addition of results from the stacking analysis. A similar figure for $f_\mathrm{H_2}$ and additional information are given in Appendix~\ref{app:PDF}. The combined analysis indicates that the high-S/N QPSB sample misses a substantial fraction of dusty and gas-rich QPSBs [$\log \,(M_\mathrm{H_2}/M_\odot) >  9-9.5$]. About a half of all face-on QPSBs have $f_\mathrm{H_2} \gtrsim 5\%$. This corresponds to $M_\mathrm{H_2} > 3 \times 10^{9} \,M_\odot$ for the QPSBs with $\log \,(M_\star/M_\odot) = 10.5-11$, and to $M_\mathrm{H_2} > 10^{9} \,M_\odot$ for those with $\log \,(M_\star/M_\odot) = 10-10.5$. Due to extra assumptions and complexities involved in reconstructing the gas distributions, this figure should be taken merely as a qualitative visual aid of our main results.

Taken together, the stacked and individual analyses of QPSBs spectra indicate that QPSBs have a wide range of H$\alpha$/H$\beta$ ratios and $f_\mathrm{H_2}$: H$\alpha$/H$\beta \approx 3-8$ and $f_\mathrm{H_2} \approx 1\%-20\%$, with the median $f_\mathrm{H_2} \approx 4\%-6\%$ or the median $M_\mathrm{H_2} \approx (1-3) \times 10^{9} \,M_\odot$.

\begin{deluxetable*}{lccccccc}
\tabletypesize{\footnotesize}
\tablecolumns{8} 
\tablewidth{0pt}
\tablecaption{Molecular Gas Masses and Molecular Gas Fractions\label{tbl:fgas}}
\tablehead{
\colhead{Sample} & \colhead{$\log\,(M_\star/M_\odot$)} & \colhead{H$\alpha$/H$\beta$} & \colhead{Median $\log\,(M_\mathrm{H_2}/M_\odot)$} & \colhead{Median $\log\,f_\mathrm{H_2}$} & \colhead{$q=0.85 \,\log\,f_\mathrm{H_2}$} & \colhead{$q=0.15 \,\log\,f_\mathrm{H_2}$} & \colhead{$N$ Galaxies} \\
\colhead{(1)} &
\colhead{(2)} &
\colhead{(3)} &
\colhead{(4)} &
\colhead{(5)} &
\colhead{(6)} &
\colhead{(7)} &
\colhead{(8)} 
}
\startdata
QPSBs  & $10-10.5$ & $3.3\,(2.1, 4.5)$ & $8.7\,(8.5, 9.1)$ & $-1.6\,(-1.8, -1.2)$ & $-1.2\,(-1.4, -0.8)$ & $-1.9\,(-2.0, -1.6)$ & ""125 \\ 
QPSBs  & $10.5-11$ & $3.3\,(2.6, 4.7)$ & $8.9\,(8.7, 9.3)$ & $-1.8\,(-2.0, -1.4)$ & $-1.5\,(-1.7, -1.1)$ & $-2.1\,(-2.3, -1.8)$ & """79 \\ 
Seyferts & $10-10.5$ & $4.3\,(3.7, 5.4)$ & $9.0\,(8.8, 9.3)$ & $-1.3\,(-1.5, -1.1)$ & $-0.9\,(-1.1, -0.7)$ & $-1.7\,(-1.8, -1.5)$ & ""105 \\ 
Seyferts & $10.5-11$ & $4.7\,(4.1, 6.2)$ & $9.3\,(9.1, 9.6)$ & $-1.4\,(-1.6, -1.1)$ & $-1.1\,(-1.3, -0.8)$ & $-1.9\,(-2.0, -1.7)$ & ""212 \\ 
QGs & $10-10.5$ & $2.7\,(1.6, 3.7)$ &$8.6\,(8.5, 8.9)$ & $-1.7\,(-1.8, -1.5)$ & $-1.3\,(-1.5, -1.1)$ & $-2.0\,(-2.1, -1.8)$ & "5601 \\ 
QGs  & $10.5-11$ & $2.8\,(1.7, 3.9)$ &$8.9\,(8.7, 9.1)$ & $-1.9\,(-2.1, -1.6)$ & $-1.6\,(-1.8, -1.4)$ & $-2.2\,(-2.4, -2.0)$ & 17121 \\ 
SFGs   & $10-10.5$ & $4.1\,(3.7, 4.7)$ & $9.1\,(8.9, 9.3)$ & $-1.1\,(-1.3, -1.0)$ & $-0.9\,(-1.0, -0.7)$ & $-1.5\,(-1.7, -1.4)$ & 37940 \\ 
SFGs & $10.5-11$ & $4.6\,(3.9, 5.4)$ & $9.4\,(9.2, 9.6)$ & $-1.3\,(-1.5, -1.1)$ & $-1.0\,(-1.3, -0.8)$ & $-1.8\,(-1.9, -1.6)$ & 31706 \\ 
\enddata 
\tablecomments{This table is based on individual galaxies with well-measured ($>3\,\sigma$) H$\alpha$/H$\beta$ ratios. We use the notation $X\,(Y, Z)$ to denote $X =$ median (50\%), $Y = 15\%$, and $Z= 85\%$ of a distribution.}
\end{deluxetable*}

\begin{deluxetable*}{lccccccc}
\tabletypesize{\footnotesize}
\tablecolumns{8} 
\tablewidth{0pt}
\tablecaption{Spectral Stacking Analysis of Low-S/N QPSBs \label{tbl:stackLSN}}
\tablehead{
\colhead{ID} & \colhead{$\log\,(M_\star/M_\odot$)} &  \colhead{Stack H$\alpha$/H$\beta$}  & \colhead{Stack $\log\,(M_\mathrm{H_2}/M_\odot)$} & \colhead{Stack $\log\,f_\mathrm{H_2}$} &  \colhead{H$\alpha$/H$\beta$} & \colhead{$\log\,(M_\mathrm{H_2}/M_\odot)$} & \colhead{$N$ Galaxies} \\
\colhead{(1)} &
\colhead{(2)} &
\colhead{(3)} &
\colhead{(4)} &
\colhead{(5)} &
\colhead{(6)} &
\colhead{(7)} &
\colhead{(8)}
}
\startdata
3  & $10.5 -11$ & $8.0 \pm 1.1$ & $10.0\,(9.8, 10.2)$  & $-0.7\,(-0.9, -0.5)$ & -- & -- & 168 \\  
3a  & $10.5 -11$ & $8.1 \pm 1.3$ & $10.0\,(9.7, 10.3)$  & $-0.7\,(-1.0, -0.5)$ & $7.8\,(6.1, 10.9)$ & $10.0\,(9.7, 10.4)$ &"47 \\  
3b & $10.5 -11$ & $3.6 \pm 0.4$ & $9.0\,(8.8, 9.2)$  & $-1.7\,(-1.8, -1.5)$ & $3.8\,(2.3, 5.0)$ & $9.1\,(8.8, 9.4)$ &"61 \\  
3c & $10.5 -11$ & $>5.6$ & $>9.5$  & $> -1.3$ & -- & -- &"65 \\  
\hline
4  & $10 -10.5$ & $5.0 \pm 0.3$ & $9.3\,(9.1, 9.4)$  & $-1.1\,(-1.2, -0.9)$ & -- & -- & 159 \\  
4a & $10 -10.5$ & $5.4 \pm 0.7$ & $9.3\,(9.1, 9.5)$  & $-0.9\,(-1.2, -0.8)$ & 7.2\,(5.4, 8.7)  & $9.7\,(9.3, 10)$ &"36 \\  
4b & $10 -10.5$ & $6.0 \pm 1.1$ & $9.5\,(9.2, 9.7)$  & $-0.8\,(-1.1, -0.6)$ & $3.9\,(3.5, 4.8)$ & $9.0\,(8.8, 9.2)$ &"39 \\  
4c  & $10 -10.5$ & $2.3 \pm 0.4$ & $8.6\,(8.5, 8.7)$  & $-1.75\,(-1.8, -1.7)$ & $2.1\,(1.5, 2.6)$ & $8.6\,(8.5, 8.7)$ &"31 \\  
4d  & $10 -10.5$ & $>4.8$ & $>9.1$  & $> -1.2$ & -- & -- &"53 \\ 
\enddata 
\tablecomments{The two samples without letter identification include all galaxies in their corresponding stellar mass ranges. Galaxies in subsamples 3c and 4d are nondetections with S/N of H$\alpha$ or H$\beta$ emission lines $<1\,\sigma$. The galaxies in the rest of the subsamples have marginally measured ($1-3\,\sigma$) H$\alpha$ or H$\beta$ lines. Subsamples 3a and 3b are split using H$\alpha$/H$\beta \ge 5.5$ and H$\alpha$/H$\beta < 5.5$, respectively. Likewise, subsamples 4a, 4b, and 4c are split using H$\alpha$/H$\beta < 5$, H$\alpha$/H$\beta = 3 - 5$, and H$\alpha$/H$\beta \ge 5$. Columns (6) and (7) give results that are based on individual spectra. The gas estimates use the median scaling relation. The lower limits of H$\alpha$/H$\beta$ (measured from the stacked spectra) are $3\,\sigma$, namely, $\frac{{\rm H}\alpha - 3 \sigma_\mathrm{H\alpha}}{3 \sigma_\mathrm{H\beta}}$.}
\end{deluxetable*}

\begin{deluxetable*}{lcccccc}
\tabletypesize{\footnotesize}
\tablecolumns{7} 
\tablewidth{0pt}
\tablecaption{Spectral Stacking Analysis  by WISE Flux Ratio of QPSBs with Low-S/N H$\alpha$/H$\beta$ \label{tbl:stackWL}}
\tablehead{
\colhead{ID} & \colhead{$\lfwise$} & \colhead{$\log\,(M_\star/M_\odot$)} &  \colhead{Stack H$\alpha$/H$\beta$}  & \colhead{Stack $\log\,(M_\mathrm{H_2}/M_\odot)$} & \colhead{Stack $\log\,f_\mathrm{H_2}$} & \colhead{$N$ Galaxies}  \\
\colhead{(1)} &
\colhead{(2)} &
\colhead{(3)} &
\colhead{(4)} &
\colhead{(5)} &
\colhead{(6)} &
\colhead{(7)}
}

\startdata
5a & $\ge 0.12$ & $10.5 -11$ & $8.1 \pm 1.3$ & $10.0\,(9.8, 10.3)$  & $-0.7\,(-1.0, -0.5)$ & 48 \\  
5b & $-0.09-0.12$   & $10.5 -11$ & $6.5 \pm 1.2$ & $9.7\,(9.4, 10)$  & $-1.0\,(-1.3, -0.7)$ & 50 \\  
5c & $<-0.09$ & $10.5 -11$ & $6.0 \pm 1.5$ & $9.6\,(9.2, 10)$  & $-1.1\,(-1.5, -0.7)$ & 49  \\ 
5d & $ < 2\,\sigma$ & $10.5 - 11$ & $4.6 \pm 1.1$ & $9.3\,(8.9, 9.7)$ & $-1.3\,(-1.7,-1.0)$ & 25 \\ 
\hline
6a & $\ge 0.17$  & $10 - 10.5$ & $4.7 \pm 0.6$ & $9.2\,(9.0, 9.4)$  & $-1.1\,(-1.3, -0.9)$ & 36 \\  
6b & $-0.01-0.17$  & $10 - 10.5$ & $>6.4$ & $>9.4$  & $>-0.8$ &  38\\ 
6c & $ < -0.01$  & $10 - 10.5$ & $6.0 \pm 1.7$ & $9.5\,(9.0, 9.8)$  & $-0.9\,(-1.4, -0.5)$ & 35 \\  
6d & $ < 2\,\sigma$ & $10 - 10.5$ & $4.6 \pm 1.0$ & $9.2\,(8.8, 9.5)$  & $-1.1\,(-1.5, -0.8)$ & 44 \\ 
\enddata 
\tablecomments{Galaxies in the subsamples with label ``d'' are nondetections in WISE. The lower limit of H$\alpha$/H$\beta$ for 6b is $3\,\sigma$.}
\end{deluxetable*}

\section{DISCUSSION}\label{sec:disc}

Understanding how and why the gas content of galaxies evolves will help reveal why galaxies stop forming stars. To that end, several studies, despite their small sample sizes, have attempted to quantify the gas and dust content of QPSBs \citep{French+15, Smercina+18} and their possible precursors \citep{Rowlands+15, Alatalo+16b, Yesuf+17, Li+19}. These studies have shown that PSBs have significant molecular gas and dust reservoirs, even when they are quiescent. As starbursts age, the molecular gas, SFR surface density, effective dust temperature, and dust-to-stellar mass ratio decrease on average \citep{Rowlands+15, Yesuf+17, French+18, Li+19}. The dust emission and masses are also broadly correlated with the molecular gas masses \citep{Yesuf+17, Li+19}.

Using dust absorption as a cost-effective probe for a much larger sample ($N \approx 530$), we confirm the previous result that large reservoirs of molecular gas are present in significant numbers of QPSBs. About a half of all face-on QPSBs have $f_\mathrm{H_2} \gtrsim 5$\%. Figure~\ref{fig:Mh2_stack} overplots the QPSBs with CO data in the same mass and redshift ranges as our sample. The ranges of $M_\mathrm{H_2}$ estimated from CO for these galaxies and those estimated by our method for the QPSBs in this work have a significant overlap. But the figure hints that the small CO sample may not be fully representative of the $M_\mathrm{H_2}$ probability density functions inferred by our method.

The progenitors of QPSBs are likely more dusty and gas-rich \citep[][]{Yesuf+14, Alatalo+16a, Alatalo+16b, Yesuf+17}. Here we also find that young and early-type Seyferts ($N \approx 320$), of which some are plausibly either the precursors or the contemporaries of QPSBs, have statistically higher gas masses. \citet{Alatalo+16b} studied CO\,(1--0) properties of 50 WISE $22\,\mu$m-detected PSBs with signatures of shocks or AGNs. The average $f_\mathrm{H_2}$ of the PSBs in this sample is at least 2 times higher (median $f_\mathrm{H_2} = 17$\%) than those of the comparison sample of normal SFGs and QPSBs (their Figure 5). The PSBs in \citet{Alatalo+16b} also have generally younger post-burst ages than QPSBs, and may be progenitors of QPSBs \citep{French+18}. Furthermore, \citet{Yesuf+17} reanalyzed \citet{Alatalo+16b}'s sample with an additional 24 low-SFR (green-valley) PSBs. The combined samples of PSBs with Seyfert-like emission-line ratios have a broad distribution of gas fraction ($f_\mathrm{H_2} = 3\%-30$\%), with mean $f_\mathrm{H_2} \approx 9$\%. In comparison, here we find that Seyferts with strong H$\delta_A$ and early-type morphologies have a median $f_\mathrm{H_2} \approx 5$\%; only about $\sim 15$\% have $f_\mathrm{H_2} > 10$\%, and $ \sim 15$\% have $f_\mathrm{H_2} < 2$\%.

In contrast to the several hundred PSBs studied by dust absorption, the PSBs studied by far-infrared emission measurements are still small. \citet{Smercina+18}, for example, studied the infrared properties of 33 QPSBs with CO observations \citep{French+15}. More than a third of these galaxies have compact ($\sim 3$\,kpc) and warm (infrared peak at $\sim 70 - 75\,\mu$m) dust reservoirs with high abundances of polycyclic aromatic hydrocarbons. \citet{Li+19} extended the analysis of \citet{Smercina+18} by including an additional 23 younger PSBs from \citet{Rowlands+15} and \citet{Alatalo+16b} that have far-infrared data. \citet{Li+19} reported a significant inverse correlation (Spearman $\rho = -0.4$), albeit with large scatter, between the specific dust mass ($M_{\rm{dust}}/M_\star$) and the time elapsed since the end of the most recent starburst. The authors estimated an exponential dust depletion timescale of $\sim 200$\,Myr. They pointed out that this timescale suggests that the dust content of PSBs should be similar to that of early-type galaxies within $\sim 1-2$\,Gyr. The authors concluded that dust may be either destroyed, expelled, or rendered undetectable over $\sim 1$ Gyr after the burst, presumably by AGN feedback. Our estimate, however, indicates that only a small fraction of all face-on QPSBs are dust-free, have $f_\mathrm{H_2} \lesssim 1\%-2\%$, and, therefore, are consistent with having similar molecular gas content as quiescent galaxies \citep{Saintonge+17}.

It should be noted that the post-burst age estimated by the methodology of \cite{French+18} indicates that QPSBs have a wide range of post-burst age\footnote{The post-burst age is defined as the time elapsed since 90\% of the stars formed in the recent burst.} ($\sim 0.1-1.0$ Gyr). We do not find a clear indication that the young QPSBs in the \citet{French+18} sample have higher H$\alpha$/H$\beta$ and $f_\mathrm{H_2}$ than do the old QPSBs in the same sample (see Appendix~\ref{app:PAge} for details). There is, however, a weak but statistically significant inverse correlation ($\rho \lesssim -0.3$) between $f_\mathrm{H_2}$ and the post-burst age when the shocked/AGN PSBs are combined with the QPSBs that have well-measured H$\alpha$/H$\beta$ ratios. Although this mean trend agrees with previous studies \citep{Yesuf+17, French+18, Li+19}, because of the large scatter, we caution against concluding that the dust and gas in (all) PSBs are declining rapidly in a coherent fashion with the post-burst age. Taken at face value, our finding that about half of the Seyfert galaxies with H$\delta_A > 4$\,{\AA} and early-type morphology have $M_\mathrm{H_2} \gtrsim 10^{9} \,M_\odot$ does not support gas removal or destruction by AGN feedback. However, ALMA observations of HCN\,(1--0) and HCO$^{+}$\,(1--0) in two CO-luminous QPSBs \citep{French+18} indicate that there is a dearth in dense gas relative to the CO-traced molecular gas, suggesting unusual conditions in which some mechanism inhibits collapse to denser states. Future observations of dense gas in more PSBs are needed to better constrain the effect of AGNs on the interstellar media of their host galaxies.

To date, there is no direct, compelling evidence that, in most local PSBs, the global gas and dust reservoirs are expelled by stellar winds or AGN feedback \citep{French+15, Rowlands+15}. While there are some individual galaxies that may have experienced AGN-driven gas outflows \citep{Baron+18}, the observed galactic winds in most local PSBs are likely not powerful enough to sweep significant amounts of gas out of their halos \citep{Yesuf+20}. Nevertheless, the average evolutionary trends of molecular gas and dust content with post-burst age, despite their large scatter, might be an indication that some PSBs do experience rapid gas and dust depletion due to AGN feedback \citep{Yesuf+17, French+18, Li+19}. Furthermore, galactic-scale outflows may play a significant role in cold gas expulsion in high-redshift PSBs \citep{Tremonti+07, Coil+11, Maltby+19}. Some observations indicate that powerful outflows appear to be much more prevalent in high-redshift, compact PSBs \citep[][]{Tremonti+07, Maltby+19}. These PSBs do not show signatures of strong AGNs in their optical spectra, suggesting either that the outflows were driven by stellar feedback, or that if AGN feedback is responsible the episodes of strong AGNs are short-lived \citep{Sell+14, Maltby+19}.

As mentioned in the Introduction, according to some cosmological simulations, AGNs are not the dominant cause of gas removal in PSBs \citep{Wild+09, Snyder+11,Davis+19}, and not all simulated PSBs are gas-deficient. In particular, \citet{Davis+19} found that local simulated PSBs have a wide range ($10^8-10^{10}\,M_\odot$) of star-forming gas masses, with half of them having $> 2 \times 10^9\,M_\odot$ (their Figure 7). Our estimated $M_\mathrm{H_2}$ distributions of QPSBs are consistent with their result. In addition, our work harbingers the possibility in the future of using deep optical spectroscopy of local QPSBs as a cost-effective alternative to constrain molecular gas distribution predicted by cosmological simulations more quantitatively and accurately.

Lastly, we raise an entirely different possibility. It is known that highly dust-enshrouded starbursts can mimic the optical spectra of PSBs \citep[e.g.,][]{Poggianti+00,Shioya+00}. Thus, the observed weak Balmer emission lines and large H$\alpha$/H$\beta$ ratios in some of our galaxies classified as ``PSBs'' may, instead, be masquerading as highly dust-obscured, still-starbursting galaxies. Future radio and far-infrared observations are needed to improve the PSB selection and rule out this possibility.

To summarize this section: the wide range of dust absorption and inferred molecular gas mass in face-on PSBs are consistent with several previous works. We do not find a clear indication that there is a coherent evolution (decline) of dust and molecular gas with the post-burst age among QPSBs. While some QPSBs might evolve rapidly \citep{Yesuf+17, French+18, Li+19}, significant numbers of QPSBs likely have not experienced complete exhaustion or expulsion of gas and dust during their lifetimes ($\sim 1$ Gyr). Multiple episodes of starburst and AGN activity may be needed to complete the migration of these gas-rich QPSBs to the red sequence \citep{Rowlands+15}. Better understanding of the PSB sample selection and properties (e.g., SFR and age) and the scatters of the mean gas and dust evolution trends are needed to firmly establish whether or not the evolution of SFR and molecular gas in PSBs is rapid, and to test the role played by AGN feedback.

\section{SUMMARY AND CONCLUSIONS}\label{sec:conc}

This work investigates the dust absorption (H$\alpha$/H$\beta$ ratio) and molecular gas content in face-on PSB galaxies at $z = 0.02 - 0.15$ with stellar masses $\log \,(M_\star/M_\odot) = 10-11$, as well as their implications. We use an empirical calibration based on our recent work \citep{Yesuf+19}, which linearly combines nebular dust absorption, average gas-phase metallicity inferred from the stellar mass, and the half-light radius to estimate molecular gas masses. We statistically infer the molecular gas either by using PSBs with well-measured H$\alpha$/H$\beta$ ratios or/and by measuring this ratio from stacked spectra. Using our new method, this work aims to confirm previous results based on CO \citep{French+15,Rowlands+15,Alatalo+16b,Yesuf+17} or dust continuum emission \citep{Smercina+18,Li+19}. While the applicability of this method to all PSBs should be subjected to further scrutiny, its predictions are fully consistent with the existing, small sample of QPSBs with direct measurements of gas and dust emission. Our uniformly selected sample of QPSBs, more than an order of magnitude larger than previously available, also allows us to account for the weak trend of molecular gas fraction with stellar mass. Our main conclusions are as follows:

\begin{itemize}

\item QPSBs have a wide range of H$\alpha$/H$\beta$ ratios, $\sim 3-8$. Some are dust-free, like most quiescent galaxies, while a significant fraction ($\gtrsim 40\%$) of QPSBs are as dusty ($A_V \approx 1.5-3$ mag) as SFGs of similar stellar masses.

\item Our dust-absorption-based gas estimator \citep{Yesuf+19} reasonably recovers existing CO-based molecular mass estimates of PSBs. This heterogenous sample contains $\sim$100 PSBs with different selection criteria, and SFRs, and AGN properties. Nevertheless, the CO luminosities of previously studied PSBs correlate significantly (Kendall $\tau \approx 0.5$; Spearman $\rho \approx 0.6-0.7$) with the molecular gas masses predicted by our scaling relation.  Comparison of our predicted molecular gas masses with the observed CO luminosities implies a median (15\%, 85\%) $\alpha_\mathrm{CO} = 4.8 \,(1, 15)\, M_\odot\,(\mathrm{K\,km\,s^{-1}\,pc^{2}})^{-1}$. Almost all QPSBs have $\alpha_\mathrm{CO}$ values that are consistent with the Milky Way value. 

\item The 204 QPSBs that have well-measured H$\alpha$/H$\beta$ ratios ($>3\,\sigma$) exhibit a broad range of $f_\mathrm{H_2} \approx 1\%-6\%$, with a median $f_\mathrm{H_2} \approx 2\%$. Most of the QPSBs have molecular gas content similar to that of QGs, while some are as gas-rich as SFGs (Table~\ref{tbl:stackHSN}). For those QPSBs with $\log \,(M_\star/M_\odot) = 10.5-11$,  the median $M_\mathrm{H_2} \approx 10^{9} \,M_\odot$.

\item Stacking analysis of 327 QPSBs with low-S/N H$\alpha$/H$\beta$ indicates that they are on average more dusty than QPSBs with well-measured H$\alpha$/H$\beta$ (Tables~\ref{tbl:stackLSN} and \ref{tbl:stackWL}). The average H$\alpha$/H$\beta \approx 5$ ($A_V \approx 2$\,mag) for the low-mass bin, H$\alpha$/H$\beta \approx 8$ ($A_V \approx 3.6$\,mag) for the high-mass bin, and median $f_\mathrm{H_2} \gtrsim  5\%-20\%$. For example,  $\sim 67\%$ (or 112 galaxies) of the stacked QPSBs with $\log \,(M_\star/M_\odot) = 10.5-11$ have $M_\mathrm{H_2} \gtrsim (3-10) \times 10^{9} \,M_\odot$. 

\item The combined analysis of individual and stacked spectra suggests that close to half of all QPSBs have $f_\mathrm{H_2} \gtrsim 5\%$. This corresponds to $M_\mathrm{H_2} > 3 \times 10^{9} \,M_\odot$ for the QPSBs with $\log \,(M_\star/M_\odot) = 10.5-11$, and to $M_\mathrm{H_2} > 10^{9} \,M_\odot$ for those with $\log \,(M_\star/M_\odot) = 10-10.5$.

\item Seyfert galaxies ($N = 317$) with young stellar population (H$\delta_A > 4$\,{\AA}) and early-type morphology ($C > 2.6$, and $\Sigma_\star > 10^{8.5}\,M_\odot$\,kpc$^{-2}$) have a median $f_\mathrm{H_2} \approx 4\%-5\%$. This corresponds to a median $M_\mathrm{H_2} \approx (1-2) \times 10^{9} \,M_\odot$. Their 0.15 or 0.85 quantiles have $f_\mathrm{H_2}$ as low as $\sim 1\%-2\%$ or as high as $\sim 8\%-25\%$. Thus, large reservoirs of molecular gas are present in significant numbers of young galaxies with active black holes. The global interstellar medium of these galaxies evidently has not been blown out by AGN feedback.

\end{itemize}

Our results add to the accumulating evidence that strong AGNs are not necessarily gas-deficient \citep[e.g.,][]{Ho+08,Husemann+17,Rosario+18,Shangguan+18, Ellison+18, ShangguanHo2019, ZhuangHo20}, and that they seemingly are not efficient at completely clearing gas out of galaxies rapidly ($\lesssim 1$\,Gyr). Better measurements of star formation, gas content, and AGN activity of PSBs are vital to put firm constraints on their star formation efficiency, thereby more accurately testing feedback models. Future studies that explain the cause of diversity in the gas fractions of PSBs will also be useful.

\medskip 

\acknowledgements

We are very grateful to the anonymous referee for helpful feedback, comments, and suggestions that significantly improved the content and presentation of the paper. This work was supported by the National Science Foundation of China (11721303, 11950410492, 11991052) and the National Key R\&D Program of China (2016YFA0400702).

Funding for SDSS has been provided by the Alfred P. Sloan Foundation, the Participating Institutions, the National Science Foundation, and the U.S. Department of Energy Office of Science. The SDSS-III website is http://www.sdss3.org/.  
SDSS-III is managed by the Astrophysical Research Consortium for the Participating Institutions of the SDSS-III Collaboration including the University of Arizona, the Brazilian Participation Group, Brookhaven National Laboratory, Carnegie Mellon University, University of Florida, the French Participation Group, the German Participation Group, Harvard University, the Instituto de Astrofisica de Canarias, the Michigan State/Notre Dame/JINA Participation Group, Johns Hopkins University, Lawrence Berkeley National Laboratory, Max Planck Institute for Astrophysics, Max Planck Institute for Extraterrestrial Physics, New Mexico State University, New York University, Ohio State University, Pennsylvania State University, University of Portsmouth, Princeton University, the Spanish Participation Group, University of Tokyo, University of Utah, Vanderbilt University, University of Virginia, University of Washington, and Yale University.

\appendix

\twocolumngrid

\section{Additional Analysis of PSBs with CO Data}\label{app:COPSB}

We revisit the properties of the samples of PSBs with CO data, which are used in Section~\ref{sec:CO} to show that our method of predicting the molecular gas mass \citep{Yesuf+19} can be applied to PSBs. We show that these samples have a wide range of SFRs, spanning from those of starburst to QGs. Hence, a variable $\alpha_{\rm CO}$ is expected.  We further show that these samples exhibit a wide range of (stacked) H$\alpha$/H$\beta$ ratios, as expected from their CO luminosities.

\subsection{Details on the Properties of PSBs with CO Measurements}

\begin{figure*}
\includegraphics[width=0.48\linewidth]{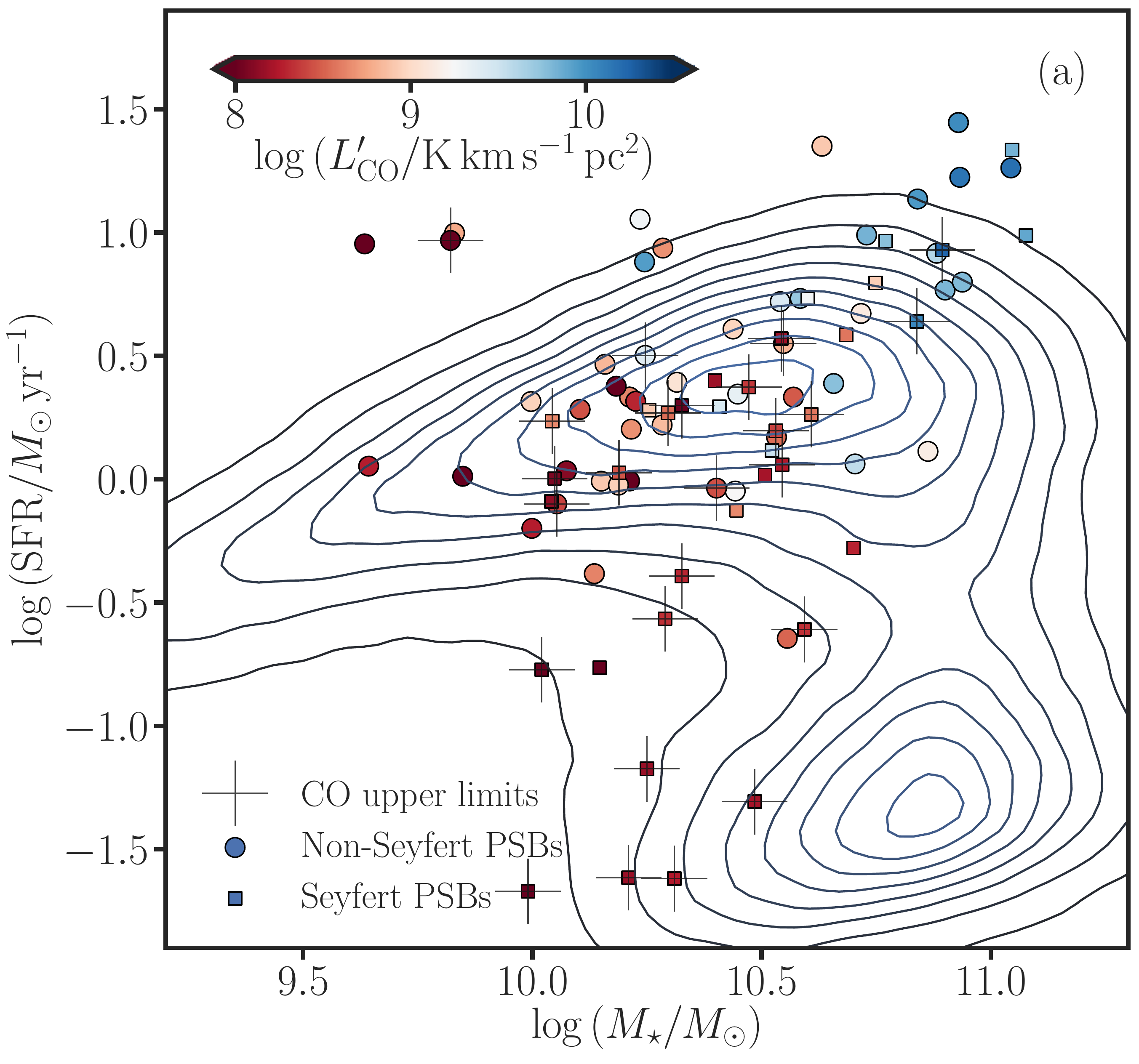}
\includegraphics[width=0.46\linewidth]{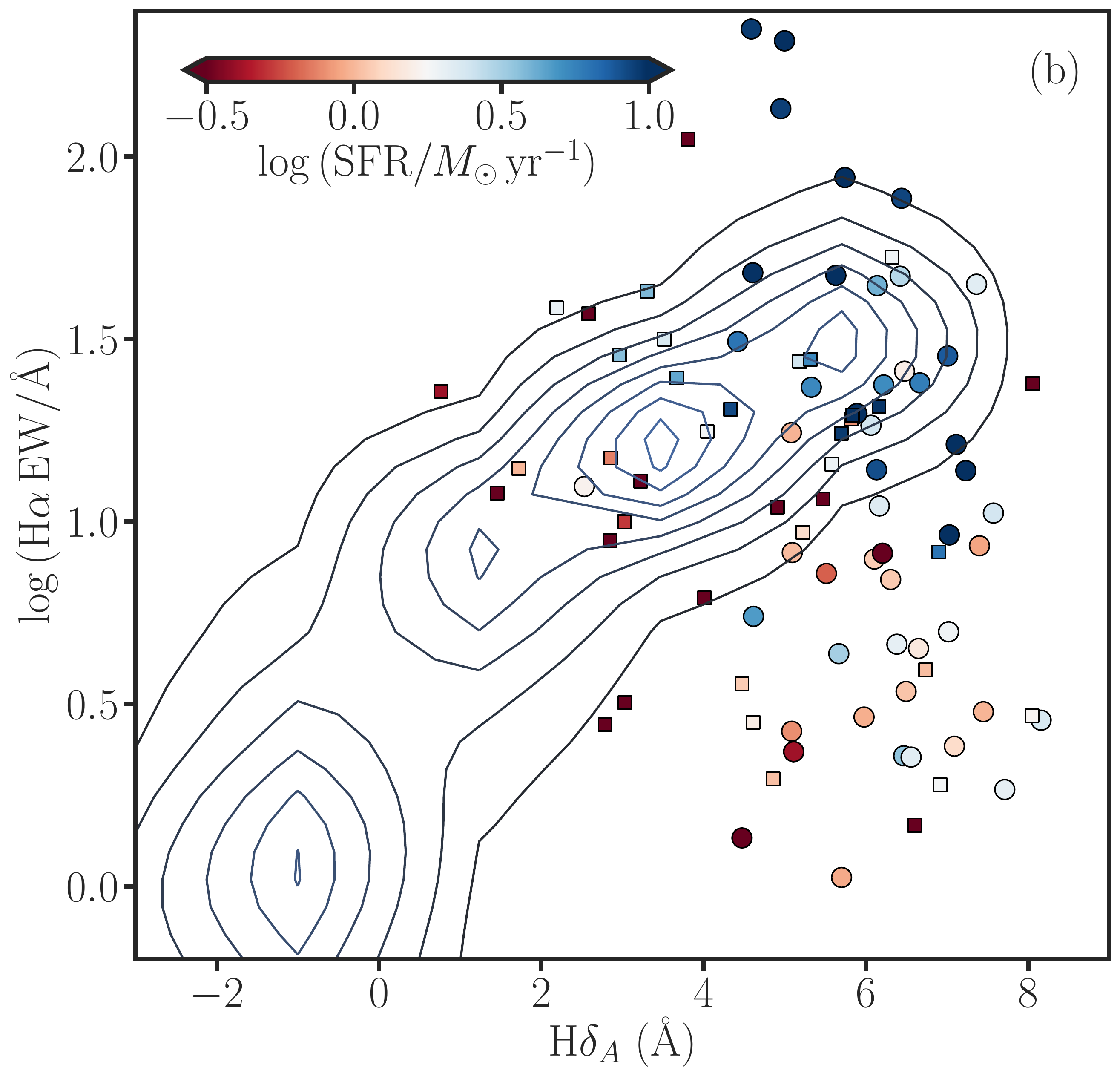}
\caption{Star formation properties of PSBs with CO data. Panel (a) shows SFR versus stellar mass, with colors indicating CO luminosity. Panel (b) shows EW of H$\alpha$ emission versus H$\delta$ absorption index, with colors indicating SFR. The SFR is estimated by spectral energy distribution fitting \citep{Salim+16, Salim+18}. The background contours represent the general galaxy population from SDSS. The PSBs exhibit a wide range of SFRs, some in the regime of starbursts. \label{fig:MS_CO}}
\end{figure*}

Figure~\ref{fig:MS_CO}a overplots PSBs with CO data on the stellar mass versus SFR diagram of galaxies at $z < 0.15$.  The SFRs are taken from version 2 of the GALEX-SDSS-WISE Legacy Catalog \citep[GSWLC-2;][]{Salim+16, Salim+18}\footnote{http://pages.iu.edu/$\sim$salims/gswlc/}. GSWLC-2 uses WISE photometry in the 12\,$\mu$m and 22\,$\mu$m bands jointly with ultraviolet and optical photometry to perform spectral energy distribution fitting to derive more accurate SFRs. PSBs span 3 orders of magnitude in SFR, ranging from quiescent to starburst galaxies.

Figures~\ref{fig:MS_CO}b plots H$\delta_A$ absorption versus H$\alpha$ EW, color-coded by SFR derived from spectral energy distribution fitting. With the exception of the Seyfert PSBs, most PSBs with strong H$\alpha$ EWs also have high SFRs, as expected. Some QPSBs have high SFRs despite their weak H$\alpha$ EWs. \cite{Salim+16} also showed that some galaxies may have weak H$\alpha$ but high SFRs, overlapping with the values of normal SFGs. Two explanations are possible.  On the one hand, differences in star formation timescales can cause H$\alpha$ to be absent during a PSB phase, while ultraviolet and mid-infrared emission still persist after the O stars have died off.  On the other hand, spatial gradients in star formation could be such that the SDSS fiber may miss off-centered regions of star formation.

\begin{figure}
\includegraphics[width=0.98\linewidth]{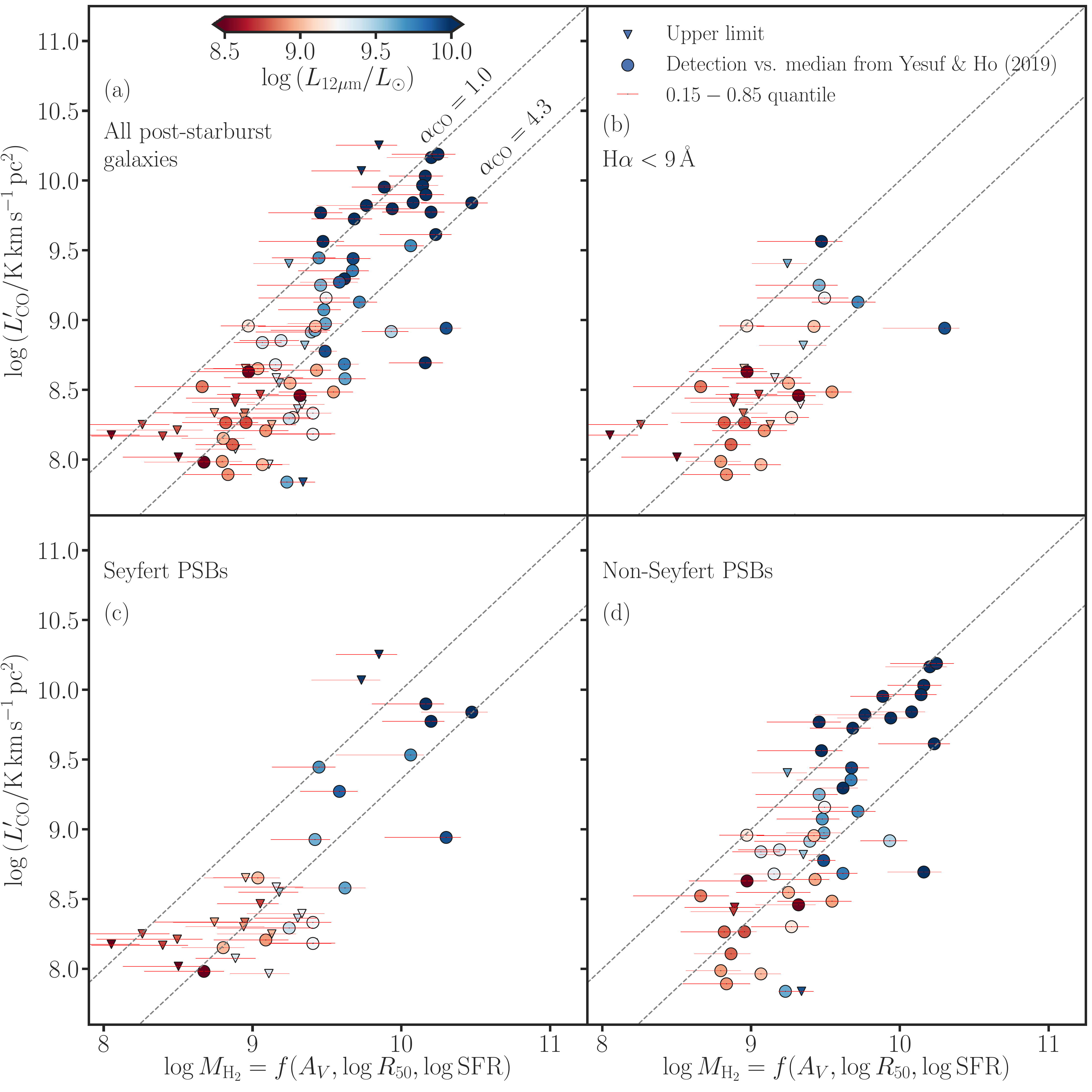}
\caption{Similar to Figure~\ref{fig:AV_CO}. Here the gas scaling relation from \cite{Yesuf+19} uses SFR from spectral energy distribution fitting, and the color-coding is by WISE 12\,$\mu$m luminosity, which is also correlated with SFR \citep{Cluver+17} and CO luminosity \citep{Gao+19}. The relation used here is more accurate and has a root mean square deviation of $\lesssim 0.3$ dex. This figure suggests that the variation in $\alpha_{\rm CO}$ likely makes a significant contribution to the scatter in Figure~\ref{fig:AV_CO}, and that most PSBs are consistent with having $\alpha_{\rm CO} \approx 4.3\,M_\odot\,(\mathrm{K\,km\,s^{-1}\,pc^{2}})^{-1}$, while those with high, starburst-like SFRs are more consistent with $\alpha_{\rm CO} \approx 1\,M_\odot\,(\mathrm{K\,km\,s^{-1}\,pc^{2}})^{-1}$. This figure is included only for illustrative purposes. Because molecular gas and SFR may not necessarily track each other well in PSBs, and the SFR of PSBs may not be very accurate, we do not use this scaling relation in the main sections of the paper. \label{fig:AV_CO2}}
\end{figure}

It is customary to use SFR as a predictor of gas content \citep[e.g.,][]{Saintonge+17, Catinella+18}. \citet{Yesuf+19} showed that $M_\mathrm{H_2}$ can be predicted to within a factor of $\lesssim 2$, if SFR is used with $A_V$ and $R_{50}$. Figure~\ref{fig:AV_CO2} compares $M_\mathrm{H_2}$ predicted by this more accurate predictor and the observed CO luminosities of the PSBs. Because molecular gas and SFR may not necessarily track each other faithfully in PSBs \citep{French+15, Li+19}, and we want to study the molecular gas independent of SFR,  we do not use this scaling relation in the main sections of the paper. The aim of Figure~\ref{fig:AV_CO2} is to show that the variation in $\alpha_{\rm CO}$ likely has a significant contribution to the scatter in Figure~\ref{fig:AV_CO}, and that most PSBs are consistent with having $\alpha_{\rm CO} \approx 4.3\, M_\odot\,(\mathrm{K\,km\,s^{-1}\,pc^{2}})^{-1}$, and those that have high SFRs (starburst-like) are more consistent with $\alpha_{\rm CO} \approx 1\, M_\odot\,(\mathrm{K\,km\,s^{-1}\,pc^{2}})^{-1}$.

\subsection{Stacking Analysis of PSBs with CO Measurements}

We perform spectral stacking analysis of PSBs with CO data \citep{French+15, Alatalo+16b, Yesuf+17}. We show that these galaxies have diverse levels of dust absorption, as expected from their wide-ranging CO luminosities, gas fractions, and mid-infrared emission \citep[Figure 8 of][]{Yesuf+17}.

\begin{figure}
\centering
\includegraphics[width=0.95\linewidth]{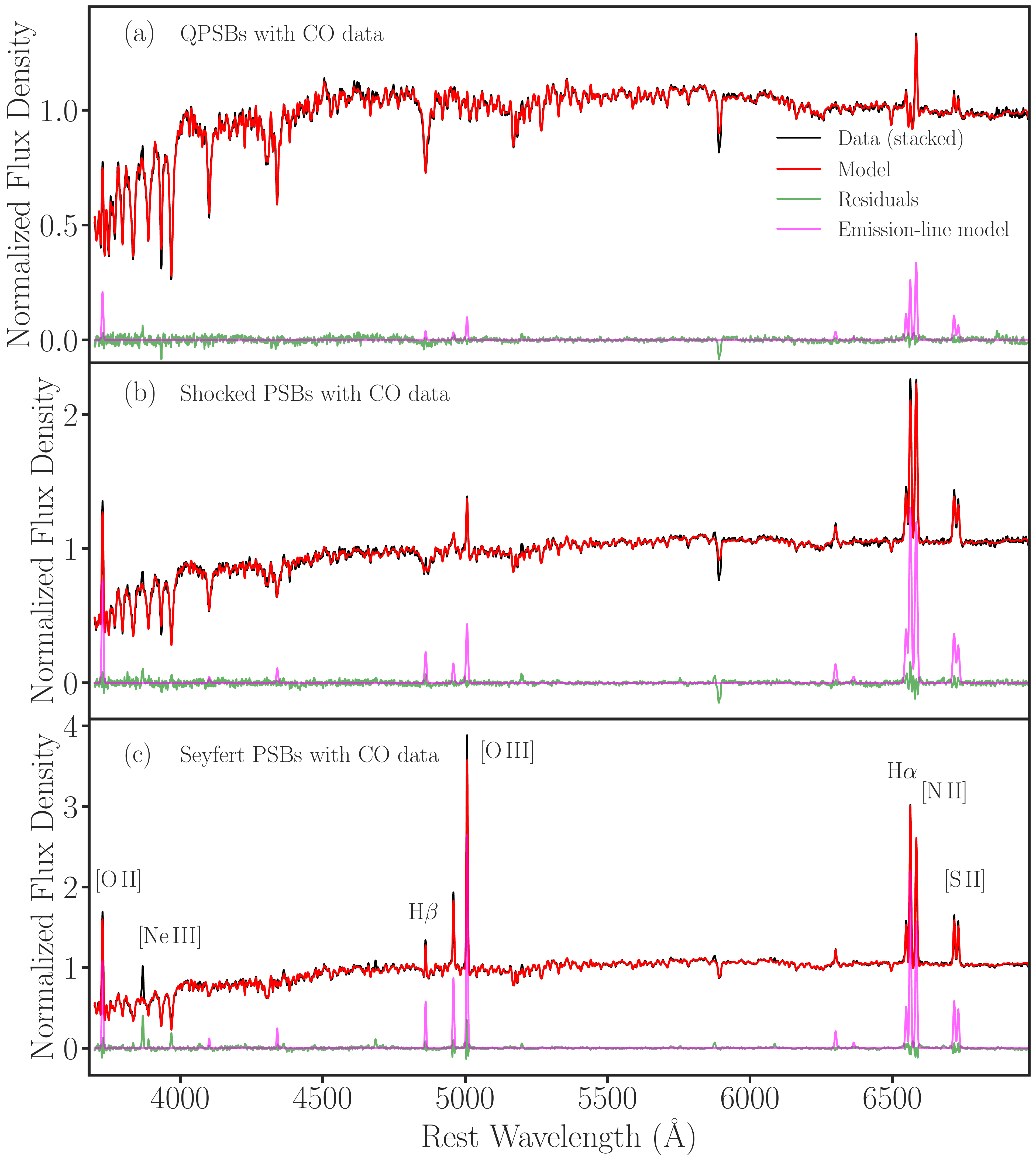}
\caption{Stacked spectra of (a) QPSBs, (b) shocked PSBs, and (c) Seyfert PSBs  with CO data \citep{French+15, Alatalo+16b, Yesuf+17}.\label{fig:PSBwCO}}
\end{figure}

Figure~\ref{fig:PSBwCO}a shows the stacked spectrum of the 32 QPSBs from the sample of \cite{French+15}. After fitting the spectra with stellar population and nebular emission models, we obtain ${\rm H}\alpha/{\rm H}\beta =  6.2 \pm 0.7$ ($A_V = 2.7 \pm 0.4$\,mag). Incidentally, previous works have adopted 3 times lower mean {\AV} for the low-S/N QPSBs in this sample \citep{French+15, Li+19}. This likely will significantly change the estimates of their dust-corrected SFRs. Figure~\ref{fig:PSBwCO}b shows the stacked spectrum and the stellar population plus emission-line model fit of the 50 PSBs with CO data from \citet{Alatalo+16b}, which the authors have highlighted as having evidence for shock ionization. This subset of PSBs has average ${\rm H}\alpha/{\rm H}\beta = 5.5 \pm 0.4$ ($A_V = 2.3 \pm 0.3$\,mag). Lastly, the 23 green-valley Seyfert PSBs (Figure~\ref{fig:PSBwCO}c) from \citet{Yesuf+17} have ${\rm H}\alpha/{\rm H}\beta = 3.5 \pm 0.4$ ($A_V = 0.4 \pm 0.4$\,mag).

\section{WISE Flux Ratio $\fwise$ as a Probe of Gas Fraction}\label{app:WISE}

We present the distribution of the WISE flux density ratio of 12\,$\mu$m to 4.6\,$\mu$m, $\fwise$, of QPSBs and discuss its relation with $f_\mathrm{H_2}$ or H$\alpha$/H$\beta$ ratios. We also show that the stacking analysis of the high-S/N QPSB sample now subdivided by $\fwise$ results in H$\alpha$/H$\beta$ and $f_\mathrm{H_2}$ distributions that are consistent with those inferred from individual spectra, similar to what we showed in Section~\ref{sec:stacking}.

\begin{figure*}
\includegraphics[width=0.48\linewidth]{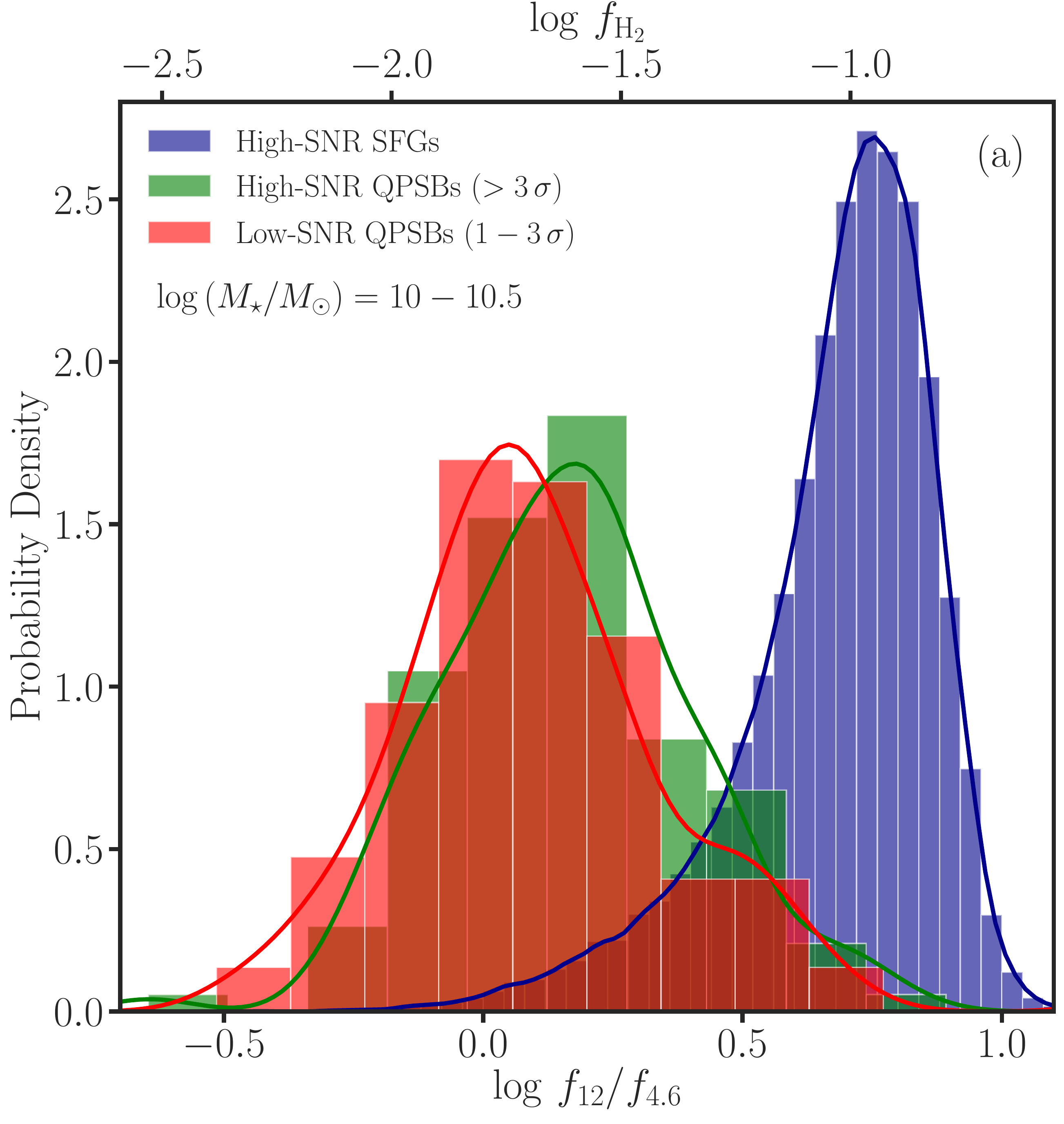}
\includegraphics[width=0.48\linewidth]{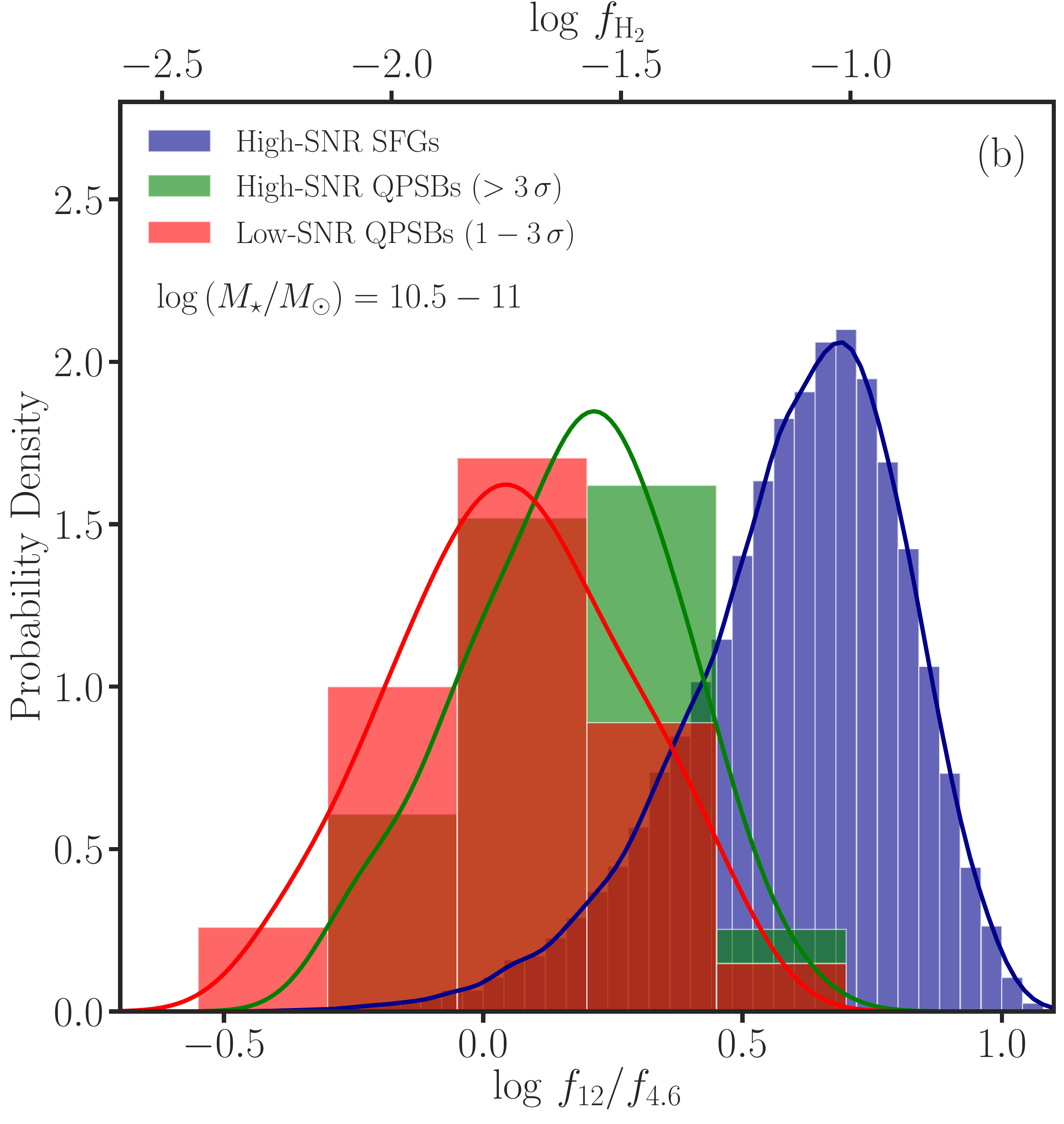}
\caption{The distribution of WISE flux density ratio 12\,$\mu$m to 4.6\,$\mu$m of face-on QPSBs and the comparison samples of SFGs in two stellar mass ranges. QPSBs have a wide range of $\fwise$ ratios, with a substantial overlap with the ratios of SFGs. The top axes show the molecular gas mass fraction $f_\mathrm{H_2}$, estimated from $\fwise$ using an empirical relation for the COLD GASS galaxies \citep{Yesuf+17}.\label{fig:w2w3}}
\end{figure*}

Figure~\ref{fig:w2w3} shows the distribution of $\fwise$ for QPSBs in the two mass ranges having S/N $> 2$ in both WISE bands. The flux density ratio $\fwise$ is a good average proxy for gas fraction for both PSBs and ordinary galaxies \citep{Yesuf+17, French+18}. However, the correlation strength between $\fwise$ and H$\alpha$/H$\beta$ is only moderate ($\rho \approx 0.4$). The top axes of the figure show the calibration using the mean $f_\mathrm{H_2}$ relation for the COLD GASS galaxies, $\log\, f_\mathrm{H_2} = 1.13 \lfwise + 1.8$. For the QPSBs with well-measured H$\alpha$/H$\beta$, adopting the WISE-COLD GASS relation gives approximately a median (15\%, 85\%) $\log\, f_\mathrm{H_2} \approx -1.6\, (-1.9, 1.3)$ for both mass ranges. The median $f_\mathrm{H_2}$ is lower by $\sim 0.2$ dex if we instead adopt the mean relation for PSBs with CO data. This estimate is consistent with the estimates from the H$\alpha$/H$\beta$ ratio given in Table~\ref{tbl:stackHSN}. The inferred gas fractions are also broadly similar (only slightly lower) for the low-S/N sample. However, the stacked analysis of H$\alpha$/H$\beta$ in Table~\ref{tbl:stackLSN} suggests a substantially higher median $\log\,f_\mathrm{H_2} \gtrsim -1$ for most low-S/N QPSBs. Although $\fwise$ is a good average $f_\mathrm{H_2}$ indicator, due to substantial scatter in their relation, it may underestimate $f_\mathrm{H_2}$ in very dusty QPSBs. On average, $\fwise$ of QPSBs is lower than that of SFGs. But there are a significant number of QPSBs that overlap with SFGs.

For completeness, Table~\ref{tbl:stackWH} checks the stacking analysis by subdividing the high-S/N QPSB sample by $\fwise$. Again, the stacking analysis recovers well the results based on individual spectra.

\begin{deluxetable*}{lcccccccc}
\tabletypesize{\footnotesize}
\tablecolumns{9} 
\tablewidth{0pt}
\tablecaption{Spectral Stacking Analysis by WISE Flux Ratio of High-S/N QPSBs\label{tbl:stackWH}}
\tablehead{
\colhead{ID} &\colhead{$\lfwise$} &  \colhead{$\log\,(M_\star/M_\odot$)} &  \colhead{Stack H$\alpha$/H$\beta$}  & \colhead{Stack $\log\,(M_\mathrm{H_2}/M_\odot)$} & \colhead{Stack $\log\,f_\mathrm{H_2}$} & \colhead{H$\alpha$/H$\beta$}  & \colhead{$\log\,(M_\mathrm{H_2}/M_\odot)$} & \colhead{$N$}  \\
\colhead{(1)} &
\colhead{(2)} &
\colhead{(3)} &
\colhead{(4)} &
\colhead{(5)} &
\colhead{(6)} &
\colhead{(7)} &
\colhead{(8)} &
\colhead{(9)} 
}
\startdata
7a & $\ge 0.12$  & $10.5 -11$ & $3.9 \pm 0.3$ & $9.1\,(9.0, 9.3)$  & $-1.6\,(-1.7, -1.4)$ &$3.5\,(2.7, 4.8)$ & $8.9\,(8.7, 9.3)$ & 50\\  
7b & $< 0.12$ & $10.5 -11$ & $3.3 \pm 0.3$ & $8.9\,(8.8, 9.1)$  & $-1.8\,(-2.0, -1.7)$ &$3.3\,(2.4, 4.8)$ & $8.9\,(8.8, 9.2)$ & 25\\  
\hline
8a & $\ge 0.17$ & $10 -10.5$ & $3.7 \pm 0.3$ & $8.8\,(8.7, 9.0)$  & $-1.4\,(-1.6, -1.3)$ &$3.6\,(2.5, 4.7)$ & $8.8\,(8.6, 9.1)$ & 55\\ 
8b & $-0.1-0.17$ & $10 -10.5$ & $3.8 \pm 0.4$ & $8.9\,(8.7, 9.0)$  & $-1.4\,(-1.6, -1.2)$ &$3.2\,(2.0, 4.5)$ & $8.7\,(8.5, 9.0)$ & 39\\ 
8c & $< -0.1$ &$10 -10.5$ & $3.7 \pm 0.4$ & $8.8\,(8.7, 9.1)$  & $-1.4\,(-1.6, -1.2)$ &$3.2\,(2.0, 3.6)$ & $8.6\,(8.4, 8.9)$ & 15\\ 
8d & $ < 2\,\sigma$ & $10 -10.5$ & $2.9 \pm 0.6$ & $8.6\,(8.5, 8.8)$  & $-1.7\,(-1.8, -1.5)$ &$2.2\,(1.7, 2.9)$ & $8.5\,(8.5, 8.7)$ & 15\\ 
\enddata 
\tablecomments{Columns (4), (5), and (6) give results that are based on stacked spectra. In comparison, Columns (7) and (8) give results that are based on individual spectra. The two approaches give similar results.}
\end{deluxetable*}

\section{Details of the Reconstructed Gas Probability Density Functions }\label{app:PDF}

We provide details of the statistical information used to approximately reconstruct the $f_\mathrm{H_2}$  or $M_\mathrm{H_2}$ distributions (Figure~\ref{fig:Mh2_stack}) of QPSBs from the stacking analysis.

We fit normal and lognormal functions, parameterized by Equations~\ref{eq:norm} and \ref{eq:lnorm}, to the distributions of inferred molecular gas masses and fractions of different subsamples of QPSBs. The parameters of the fits are given in Table~\ref{tbl:lnorm}.

\begin{equation}\label{eq:norm}
F(x = \log\,f_\mathrm{H_2},\mu, \sigma) \propto  \frac {1}{\sigma} \exp \left[-\frac{(x-\mu)^2}{2\sigma^2}\right]
\end{equation}

\begin{equation}\label{eq:lnorm}
F(x = \log\,M_\mathrm{H_2},\mu^\prime, \sigma^\prime) \propto \frac {1}{x\sigma^\prime}\exp \left[\frac{-(\ln x-\mu^\prime)^2}{2\sigma^{\prime2}}\right]
\end{equation}

\begin{figure*}
\includegraphics[width=0.48\linewidth]{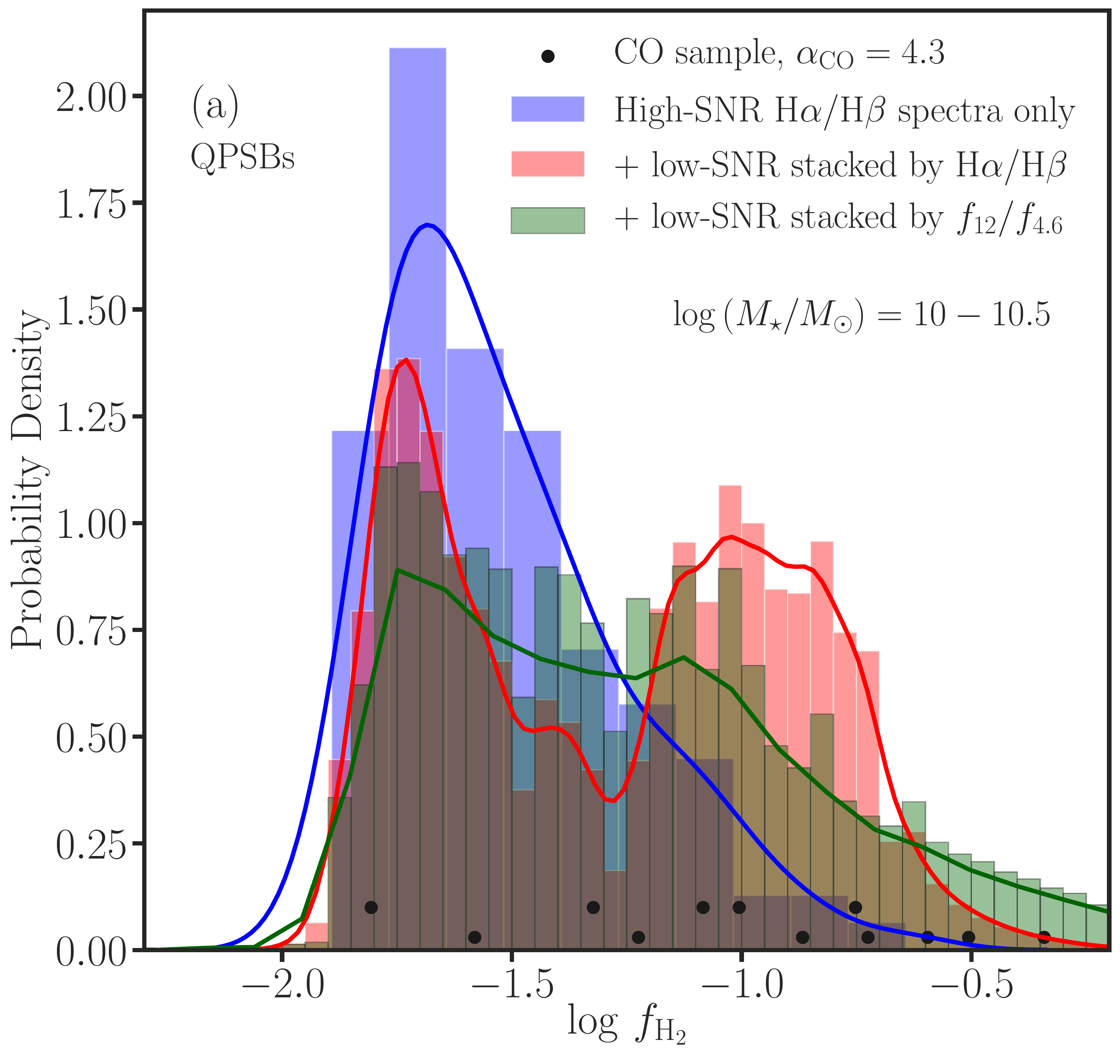}
\includegraphics[width=0.48\linewidth]{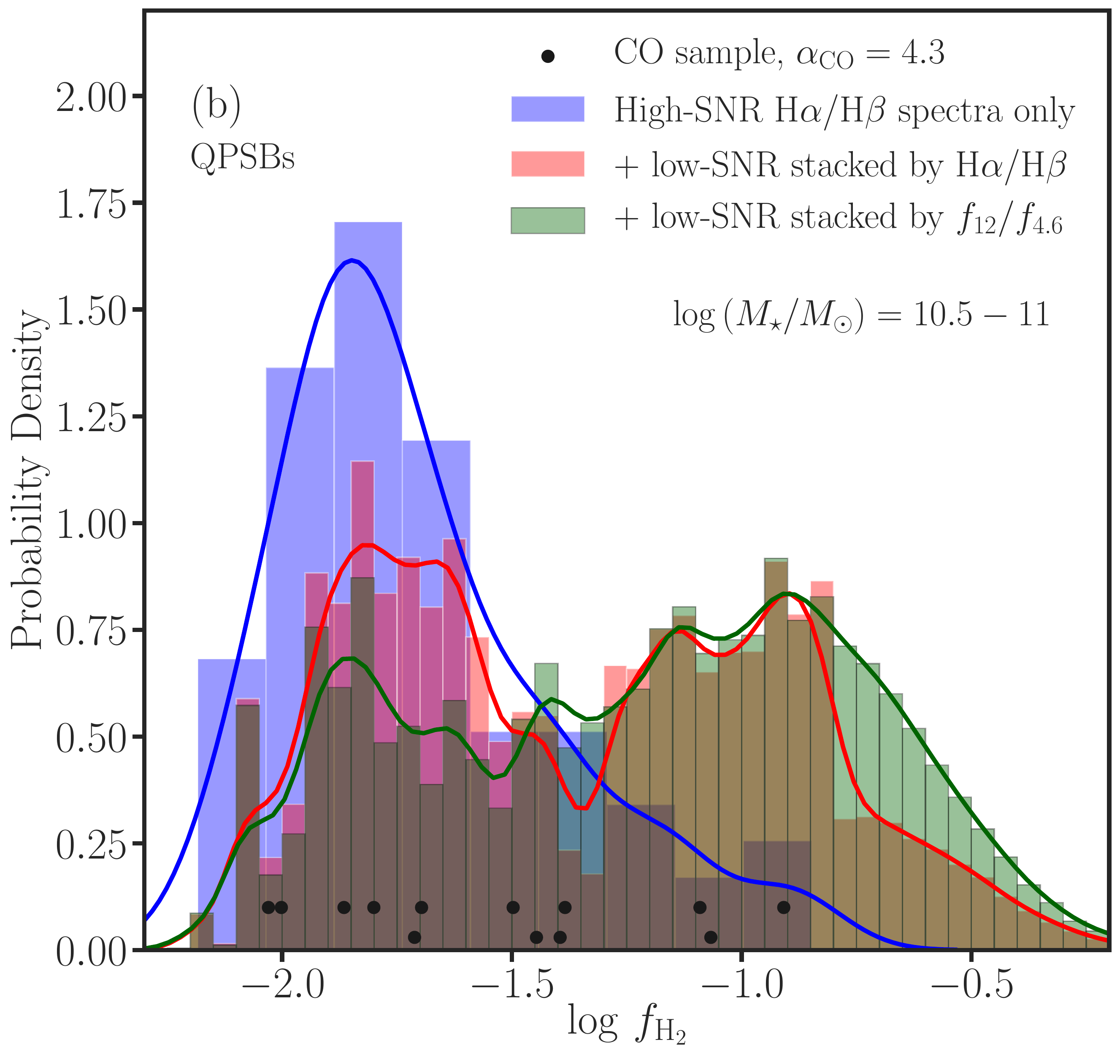}
\caption{Attempted reconstruction of molecular gas distribution of all face-on QPSBs at $z=0.02-0.15$. This figure is similar to Figure~\ref{fig:Mh2_stack}, except that here the $f_\mathrm{H_2}$ distributions are plotted instead of $M_\mathrm{H_2}$ distributions.\label{fig:fh2_stack}}
\end{figure*}

\begin{deluxetable}{ccc}
\tabletypesize{\footnotesize}
\tablecolumns{3} 
\tablewidth{0pt}
\tablecaption{Parameters of Normal and Lognormal Fits \label{tbl:lnorm}}
\tablehead{
\colhead{Sample ID} & \colhead{$\log\,M_\mathrm{H_2}$} & \colhead{$\log\,f_\mathrm{H_2}$}}
\startdata 
1a & $\mu^\prime = 2.228,\,\, \sigma^\prime = 0.017$ & $\mu = -1.41,\,\, \sigma = 0.16$\\ 
1b & $\mu^\prime = 2.212,\,\, \sigma^\prime = 0.019$ & $\mu = -1.56,\,\, \sigma = 0.18$\\ 
1c & $\mu^\prime = 2.188,\,\, \sigma^\prime = 0.018$ & $\mu = -1.79,\,\, \sigma = 0.17$\\ 
2a & $\mu^\prime = 2.226,\,\, \sigma^\prime = 0.018$ & $\mu = -1.00,\,\, \sigma = 0.17$\\ 
2b & $\mu^\prime = 2.162,\,\, \sigma^\prime = 0.016$ & $\mu = -1.55,\,\,\sigma = 0.15$\\ 
2c & $\mu^\prime = 2.149,\,\, \sigma^\prime = 0.011$ & $\mu = -1.69,\,\, \sigma = 0.10$\\
3a & $\mu^\prime = 2.303,\,\, \sigma^\prime = 0.023$ & $\mu = -0.75,\,\, \sigma = 0.24$\\ 
3b & $\mu^\prime = 2.199,\,\, \sigma^\prime = 0.017$ & $\mu = -1.66,\,\,\sigma = 0.17$\\ 
4a & $\mu^\prime = 2.234,\,\, \sigma^\prime  = 0.020$ & $\mu = -0.95,\,\, \sigma = 0.18$\\ 
4b & $\mu^\prime = 2.248,\,\, \sigma^\prime  = 0.026$ & $\mu = -0.85,\,\, \sigma = 0.25$\\ 
4c & $\mu^\prime = 2.151,\,\, \sigma^\prime  = 0.012$ & $\mu = -1.74,\,\, \sigma = 0.09$\\ 
5a & $\mu^\prime = 2.304,\,\, \sigma^\prime  = 0.022$ & $\mu = -0.72,\,\, \sigma = 0.23$\\ 
5b & $\mu^\prime = 2.274,\,\, \sigma^\prime  = 0.025$ & $\mu = -0.98,\,\, \sigma = 0.26$\\ 
5c & $\mu^\prime = 2.262,\,\, \sigma^\prime  = 0.035$ & $\mu = -1.11,\,\, \sigma = 0.35$\\ 
5d & $\mu^\prime = 2.230,\,\, \sigma^\prime  = 0.033$ & $\mu = -1.36,\,\, \sigma = 0.32$\\ 
6a & $\mu^\prime = 2.216,\,\, \sigma^\prime  = 0.020$ & $\mu = -1.13,\,\, \sigma = 0.19$\\ 
6c & $\mu^\prime = 2.242,\,\, \sigma^\prime  = 0.041$ & $\mu = -0.90,\,\, \sigma = 0.38$\\ 
6d & $\mu^\prime = 2.214,\,\, \sigma^\prime  = 0.033$ & $\mu = -1.15,\,\, \sigma = 0.30$\\ 
7a & $\mu^\prime = 2.209,\,\, \sigma^\prime = 0.015$ & $\mu = -1.58,\,\, \sigma = 0.14$\\ 
7b & $\mu^\prime = 2.188,\,\, \sigma^\prime = 0.015$ & $\mu = -1.81,\,\, \sigma = 0.14$\\ 
8a & $\mu^\prime = 2.180,\,\, \sigma^\prime = 0.015$ & $\mu = -1.41,\,\, \sigma = 0.13$\\ 
8b & $\mu^\prime = 2.182,\,\, \sigma^\prime = 0.019$ & $\mu = -1.40,\,\, \sigma = 0.17$\\ 
8c & $\mu^\prime = 2.178,\,\, \sigma^\prime = 0.019$ & $\mu = -1.43,\,\, \sigma = 0.17$\\ 
8d & $\mu^\prime = 2.154,\,\, \sigma^\prime = 0.018$ & $\mu = -1.65,\,\, \sigma = 0.15$\\ 
\enddata 
\tablecomments{The lognormal parameters can be transformed to a mean gas mass and its standard deviation as follows: \\
$\langle \log\,M_\mathrm{H_2} \rangle = \exp \mu^\prime$ and \\
$\sigma_{\log\,M_\mathrm{H_2}} = \sqrt{(\exp\,(\sigma^{\prime2})-1) \times \exp\,(2\mu^\prime+\sigma^{\prime2})}$.}
\end{deluxetable}

\section{Gas Trends with Post-burst Age}\label{app:PAge}

This appendix provides supporting analysis to the statement in the Discussion section that there is no clear indication that the young QPSBs in the \citet{French+18} sample have higher H$\alpha$/H$\beta$ and $f_\mathrm{H_2}$ than do the old QPSBs in the same sample.

\citet{French+18} attempted to accurately measure ages and burst properties of QPSBs by fitting stellar population synthesis models to ultraviolet-optical photometry and optical spectra. We use their publicly available catalog on VizieR\footnote{http://vizier.u-strasbg.fr/viz-bin/VizieR?-source=J/ApJ/862/2} to check whether there are clear trends between post-burst ages of QPSBs and their H$\alpha$/H$\beta$ ratios or their inferred molecular gas content. As in the main text, we restrict the QPSB sample to $z = 0.02 - 0.15$ and $\log \,(M_\star/M_\odot) = 10-11$. This results in 286 QPSBs. Table~\ref{tbl:page} presents the summary statistics of the distributions of H$\alpha$/H$\beta$, $M_\mathrm{H_2}$, and $f_\mathrm{H_2}$ for various subsets of QPSBs stacked by subdividing according to their post-burst ages. Although there is a weak inverse correlation ($\rho \lesssim -0.3$) between $f_\mathrm{H_2}$ and the post-burst age when the shocked/AGN PSBs are combined with the QPSBs, the stacking analysis does not indicate that the young QPSBs from \citet{French+18} have higher H$\alpha$/H$\beta$ and $f_\mathrm{H_2}$ than do the old QPSBs. The stacking analysis is consistent with no trend or, at best, a weak positive trend. We are unable to explain the diversity of gas and dust content of QPSBs as a sequence in post-burst age. Future studies should revisit this problem with better data and more thorough analysis. We note that Davis et al. (2019; their Figure 8) showed that the star-forming gas fractions of simulated PSBs decrease systematically from $\sim 30\%$ to $\sim 3\%$ for PSBs younger than $\sim 600$ Myr; the scatter increases thereafter, and aged PSBs have a wide range ($0.1\%-30$\%) of gas fractions.

\begin{deluxetable*}{lcccccc}
\tabletypesize{\footnotesize}
\tablecolumns{7} 
\tablewidth{0pt}
\tablecaption{Spectral Stacking Analysis by Post-burst Age of QPSBs\label{tbl:page}}
\tablehead{
\colhead{ID} & \colhead{Age/Myr} & \colhead{$\log\,(M_\star/M_\odot$)} &  \colhead{Stack H$\alpha$/H$\beta$}  & \colhead{Stack $\log\,(M_\mathrm{H_2}/M_\odot)$} & \colhead{Stack $\log\,f_\mathrm{H_2}$} & \colhead{$N$ Galaxies} \\
\colhead{(1)} &
\colhead{(2)} &
\colhead{(3)} &
\colhead{(4)} &
\colhead{(5)} &
\colhead{(6)} &
\colhead{(7)}
}

\startdata
9a & $> 415$  & $10.5 -11$ & $> 8.3$ & $> 9.9$  & $>-0.8$ & 42 \\  
9b& $295-415$  & $10.5 -11$ & $5.7 \pm 1.2$ & $9.6\,(9.2, 9.8)$  & $-1.2\,(-1.5, -0.9)$ & 44 \\  
9c& $< 295$  & $10.5 -11$ & $4.8 \pm 0.7$ & $9.4\,(9.2, 9.6)$  & $-1.4\,(-1.6, -1.1)$ & 42  \\ 
\hline
10a& $> 540$  & $10 - 10.5$ & $5.7 \pm 0.8$ & $9.3\,(9.1, 9.6)$  & $-0.9\,(-1.1, -0.7)$ & 50 \\  
10b& $360 - 540$  & $10 - 10.5$ & $4.1 \pm 0.6$ & $8.9\,(8.7, 9.2)$ & $-1.3\,(-1.6, -1.1)$ & 47\\  
10c & $< 360 $  & $10 - 10.5$ & $3.5 \pm 0.3$ & $8.8\,(8.6, 9.0)$ & $-1.5\,(-1.7,-1.4)$ & 61 \\ 
\hline
11a & $> 415$  & $10.5 -11$ & $4.2 \pm 1.0$ & $9.2\,(8.8, 9.5)$  & $-1.5\,(-1.9, -1.2)$ & 10 \\  
11b & $295-415$  & $10.5 -11$ & $3.3 \pm 0.9$ & $8.9\,(8.7, 9.2)$  & $-1.8\,(-2.1, -1.5)$ & "9 \\  
11c & $< 295$ & $10.5 -11$ & $3.5 \pm 0.5$ & $9.0\,(8.8, 9.2)$ & $-1.7\,(-2.0, -1.5)$ & 14  \\ 
\hline
12a & $> 540$  & $10 - 10.5$ & $5.7 \pm 1.1$ & $9.3\,(9.0, 9.6)$  & $-0.9\,(-1.2, -0.6)$ & 19 \\  
12b & $360 - 540$  & $10 - 10.5$ & $3.3 \pm 0.4$ & $8.6\,(8.5, 8.8)$ & $-1.6\,(-1.7, -1.4)$ &  24\\  
12c & $< 360 $  & $10 - 10.5$ & $3.6 \pm 0.4$ & $8.8\,(8.6, 9.0)$ & $-1.5\,(-1.7,-1.3)$ & 36 \\  
\hline
13a & $> 415$  & $10.5 -11$ & $>7.5$ & $>9.8$  & $>-0.9$ & 31 \\  
13b & $295-415$  & $10.5 -11$ & $> 6.3$ & $>9.6$  & $>-1.1$ & 33 \\  
13c & $< 295$  & $10.5 -11$ & $4.2 \pm 0.6$ & $9.2\,(9.0, 9.4)$  & $-1.5\,(-1.8, -1.3)$ & 28  \\ 
\hline
14a & $> 540$  & $10 - 10.5$ & $5.3 \pm 1.3$ & $9.3\,(8.9, 9.6)$  & $-1.0\,(-1.4, -0.7)$ & 27 \\  
14b & $360 - 540$  & $10 - 10.5$ & $3.9 \pm 1.5$ & $9.0\,(8.5, 9.4)$  & $-1.4\,(-1.8, -0.9)$ &  22\\  
14c & $< 360 $  & $10 - 10.5$ & $4.2 \pm 0.8$ & $9.1\,(8.8, 9.4)$ & $-1.3\,(-1.6, -1.0)$ & 25 \\  
\enddata 
\tablecomments{All QPSBs are stacked in subsamples 9 and 10. In subsamples 11 and 12, only those with well-measured H$\alpha$ and H$\beta$ are stacked, while in subsamples 13 and 14 only the low-S/N QPSBs are stacked.}
\end{deluxetable*}

\section{Additional Analysis of Individual High-S/N QPSBs}

This section provides additional visualization of the gas fractions of high-S/N (median continuum S/N pixel$^{-1} \, > 10$) QPSBs and the comparison samples. We also show that some PSBs have H$\alpha$/H$\beta < 2.5$, which is unphysical, likely due to spectra with still insufficient S/N (S/N pixel $^{-1}\, < 25$) or imperfect deblending of the Balmer emission and absorption lines \citep[see, e.g.,][]{Ho+97}.

\begin{figure*}
\includegraphics[scale=0.32]{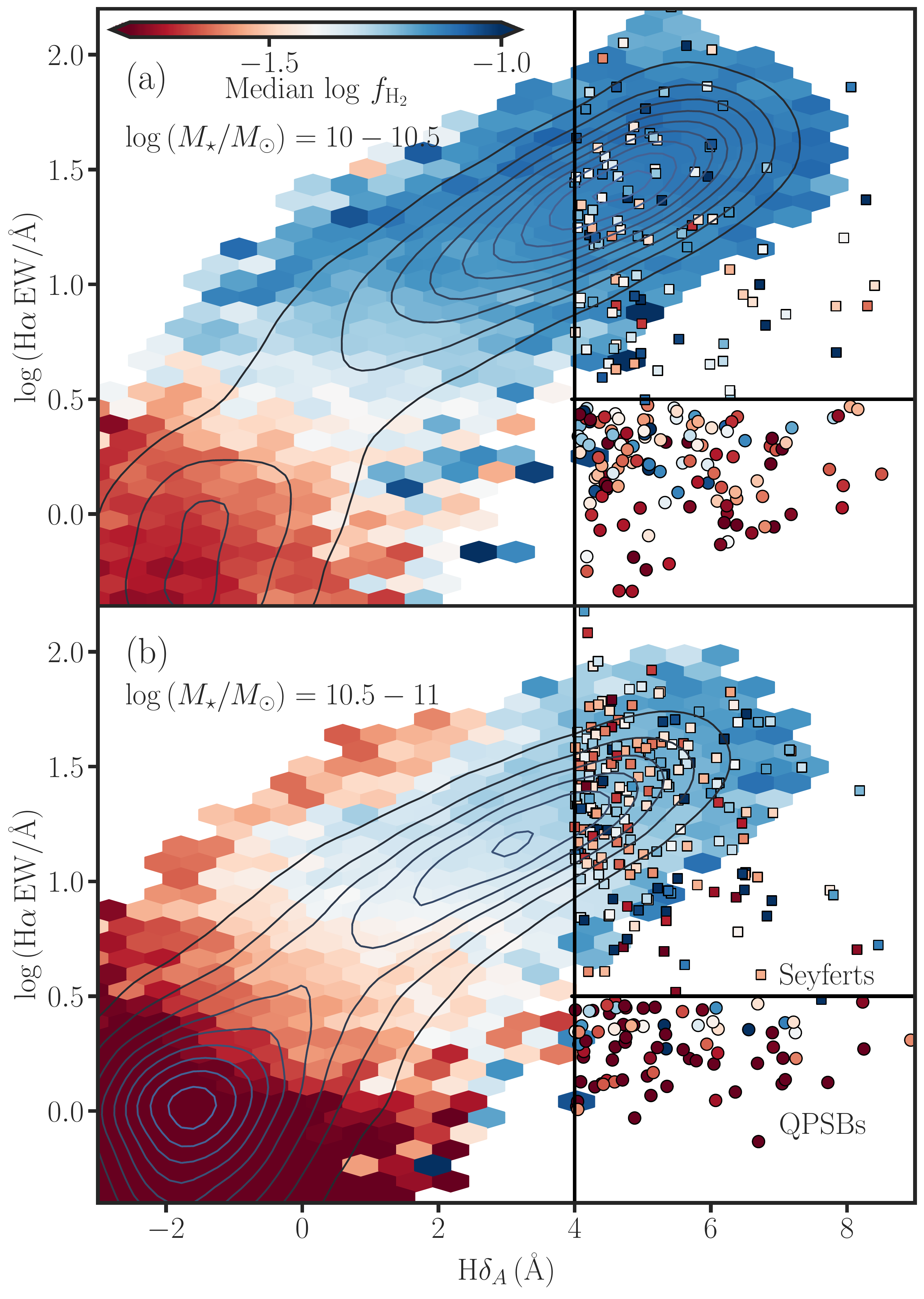} 
\includegraphics[scale=0.32]{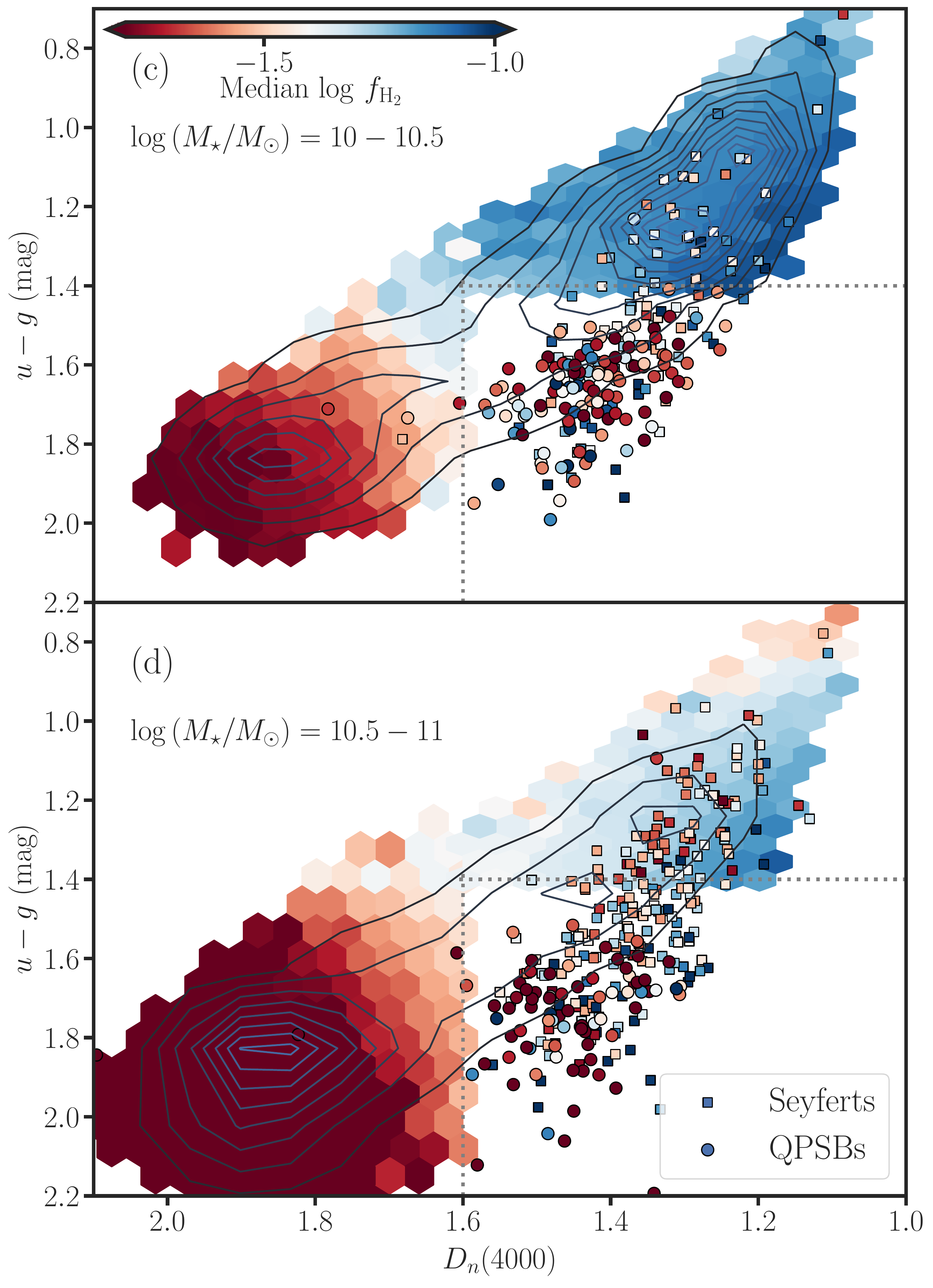}
\caption{Comparison of molecular gas fractions ($f_\mathrm{H_2}$) of QPSBs with those of SFGs and QGs. The contours show the number density of SDSS galaxies at redshifts $z = 0.02 - 0.15$ with stellar masses $\log \,(M_\star/M_\odot) = 10 - 10.5$ or $\log \,(M_\star/M_\odot) = 10.5 - 11$. The colors denote $f_\mathrm{H_2}$ (Section~\ref{sec:res}). The colored circles are high-S/N QPSBs, while the colored squares are early-type Seyferts. Panels (a) and (b) plot H$\delta_A$ absorption against H$\alpha$ emission EW. Panels (c) and (d) plot the central (fiber) 4000\,{\AA} break index versus the galaxy-wide $u-g$ color. The dotted-lines in panel (c) and (d) indicate the boundaries inside which no color scale is shown for clarity; they are not used in the sample selection of QPSBs. It is clear that PSBs have a wide range of gas fractions that overlap with those observed in SFGs or QGs. \label{fig:MH2med}}
\end{figure*}

Figures~\ref{fig:MH2med}a and \ref{fig:MH2med}b plot H$\delta_A$ absorption versus H$\alpha$ emission EW, color-coded by the median molecular gas fraction $f_\mathrm{H_2}$. The contours show the number density of all SDSS galaxies for the given stellar mass range at $z = 0.02 - 0.15$. We, however, only use galaxies with detectable H$\alpha$ and H$\beta$ lines to calculate the median $f_\mathrm{H_2}$, and we only show bins that have 10 or more galaxies. For most (nonbursty) galaxies, H$\alpha$ emission correlates with H$\delta_A$ absorption. As expected, the median $f_\mathrm{H_2}$ is also correlated with H$\alpha$ EW for normal galaxies. QPSB galaxies (colored circles), with weak emission (H$\alpha < 3$ \,{\AA}) but strong absorption (H$\delta_A > 4$\,{\AA}), deviate from the locus of normal galaxies because their star formation quenched rapidly. The colored squares are early-type Seyferts. The main point of the figures is to show that QPSBs have a broad range of $f_\mathrm{H_2}$, which overlaps with the typical gas fractions of SFGs or QGs. Some of the Seyferts also have similar $f_\mathrm{H_2}$ to SFGs. By definition, QPSBs do not contain strong AGNs. We include the young, early-type Seyferts in our analysis to strengthen our main conclusion that significant numbers of QPSBs have copious amounts of dust and gas, which probably are not removed or destroyed during their previous AGN activities. Although a recent cold gas accretion cannot be ruled out, the simpler explanation that the gas was not removed in the first place is preferred, lacking further evidence to the contrary.

Figures~\ref{fig:MH2med}c and \ref{fig:MH2med}d plot $D_n(4000)$ index versus $u-g$ color, again color-coded by the median $f_\mathrm{H_2}$. The $4000$\,{\AA} break quantifies the central mean stellar age (in the fiber), while $u-g$ color is sensitive to the galaxy-wide SFR over a timescale of several hundred Myr. Both quantities increase as the stellar population ages. Dust extinction affects $u-g$ color but not much $D_n(4000)$. The former is  corrected for the foreground Galactic extinction but not for extinction internal to the galaxies. The colored bins outside the dotted box show the sample median gas fractions in a given bin. In both mass ranges, QPSBs have intermediate colors and $D_n(4000)$ indices but a broad range of $f_\mathrm{H_2}$ that overlaps with the typical gas fractions of SFGs and QGs.

The H$\delta_A$, $D_n(4000)$, and $u-g$ color distributions of the low-S/N QPSBs are similar to those of the high-S/N QPSBs.

Finally, Table~\ref{tbl:SNmed} splits the high-S/N QPSB sample into two, and it shows that there is a correlation between the S/N of the continua and the statistical properties of H$\alpha$/H$\beta$. As the S/N increases, the 15\% quantile of H$\alpha$/H$\beta$ distribution shifts toward higher, more physical values. Likewise, the 85\% quantile of H$\alpha$/H$\beta$ distribution increases to higher values ($\sim 5-6$ depending on $M_\star$), strengthening the conclusion that some QPSBs are very dusty. Generally, plotting H$\alpha$/H$\beta$ vs. EW H$\alpha$ for all galaxies with high-S/N ($> 3$) emission lines indicates that the unphysical H$\alpha$/H$\beta$ ratios are primarily associated with weak emission lines (EW H$\alpha \lesssim 1-3$\,{\AA}). 
Therefore, in some cases, despite the high-S/N continua, the unavoidable template mismatch from the continuum subtraction will pose a limit to how accurately one can measure the weak H$\beta$ emission lines on top of strong absorption lines of some PSBs. 

\begin{deluxetable*}{lccc}
\tabletypesize{\footnotesize}
\tablecolumns{4} 
\tablewidth{0pt}
\tablecaption{Possible Systematic Effects of the Continuum S/N on H$\alpha$/H$\beta$ Measurements \label{tbl:SNmed}}
\tablehead{ \colhead{$\log\,(M_\star/M_\odot$)} &  \colhead{H$\alpha$/H$\beta$ S/N $>10$} & \colhead{H$\alpha$/H$\beta$ S/N $ = 10 - 25$} & \colhead{H$\alpha$/H$\beta$ S/N $>25$}} 
\startdata
$10-11$ & $3.3\,(2.3, 4.6)$ & $3.0\,(2.1, 4.3)$ & $3.7\,(2.7, 5.2)$ \\ 
$10.5 -11$ & $3.3\,(2.6, 4.7)$ & $3.1\,(2.4, 4.5)$ & $3.8\,(2.9, 6.2)$\\ 
$10 -10.5$ & $3.2\,(2.1, 4.5)$ & $3.0\,(2.0, 4.3)$ & $3.5\,(2.7, 4.7)$\\  
\enddata 
\end{deluxetable*}



\end{document}